\documentclass[useAMS,usenatbib,usegraphicx]{mn2e}
\usepackage{graphicx, epstopdf, txfonts, color, soul,ulem, rotating}

\def\apj{{ApJ}}                 
\def\apjl{{ApJ}}                
\def\aapr{{A\&A~Rev.}}          
\def\mnras{{MNRAS}}             
\def\solphys{{Sol.~Phys.}}      
\def\grl{{Geophysical Research Letters}}
\def\jgr{{Journal of Geophysical Research}}

\def\mn{{MNRAS}}             
\def\sp{{Sol.~Phys.}}      
\def\aanda{{A\&A}}                
\def\aandar{{A\&A~Rev.}}          

\title[]{Influence of surface stressing on stellar coronae and winds}
\author[]{M. Jardine$^1$ \thanks{E-mail: mmj@st-andrews.ac.uk}, A. A. Vidotto$^1$, A. van Ballegooijen$^2$, J.-F. Donati$^3$, J. Morin$^4$, \newauthor R. Fares,$^1$ and T.I Gombosi$^5$\\
$^1$SUPA, School of Physics \& Astronomy, University of St Andrews, North Haugh, St Andrews,  KY16 9SS, UK\\
$^2$Harvard-Smithsonian Center for Astrophysics, 60 Garden Street, Cambridge, MA02138\\
$^3$Laboratoire dÕAstrophysique, Observatoire Midi-Pyr\'en\'ees, 14 
Av. E. Belin, F-31400 Toulouse, France \\
$^4$Institut  f\"ur Astrophysik, Georg-August-Universit\"at G\"ottingen, Friedrich-Hund-Platz 1, 37077 G\"ottingen\\
$^5$University of Michigan, 1517 Space Research Building, Ann Arbor, MI, 48109-2143, USA\\
}
\begin{document}

\date{Accepted ??. Received ??; in original form ??}

\pagerange{\pageref{firstpage}--\pageref{lastpage}} \pubyear{2011}

\maketitle

\label{firstpage}

\begin{abstract}


The large-scale field of the Sun is well represented by its lowest energy (or potential) state. Recent observations, by comparison, reveal that many solar-type stars show large-scale surface magnetic fields that are highly non-potential  - that is, they have been stressed above their lowest-energy state. This non-potential component of the surface field is neglected by current stellar wind models. The aim of this paper is to determine its effect on the coronal structure and wind.
We use Zeeman-Doppler surface magnetograms of two stars - one with an almost potential, one with a non-potential surface field - to extrapolate a static model of the coronal structure for each star. We find that the stresses are carried almost exclusively in a band of uni-directional azimuthal field that is confined to mid-latitudes. Using this static solution as an initial state for an MHD wind model, we then find that the final state is determined primarily by the potential component of the surface magnetic field. The band of azimuthal field must be confined close to the stellar surface as it is not compatible with a steady-state wind. By artificially increasing the stellar rotation rate we demonstrate that the observed azimuthal fields can not be produced by the action of the wind but must be due to processes at or below the stellar surface. We conclude that the background winds of solar-like stars are largely unaffected by these highly-stressed surface fields. Nonetheless, the increased flare activity and associated coronal mass ejections that may be expected to accompany such highly-stressed fields may have a significant impact on any surrounding planets.

\end{abstract}

\section{Introduction}

The magnetic fields of solar-like stars are an important influence not only on the rotational evolution of the stars themselves, but also on the atmospheres and exospheres of any planets that might surround them.  This magnetic field not only transfers torques between the protoplanetary disk and the young star, but it also governs the loss of angular momentum in a wind. Any orbiting planets are exposed to the erosive effects of this wind and also the coronal X-ray emission from the star \citep{khodachenko_CME_07}.

\begin{figure*}
	\includegraphics[width=57mm]{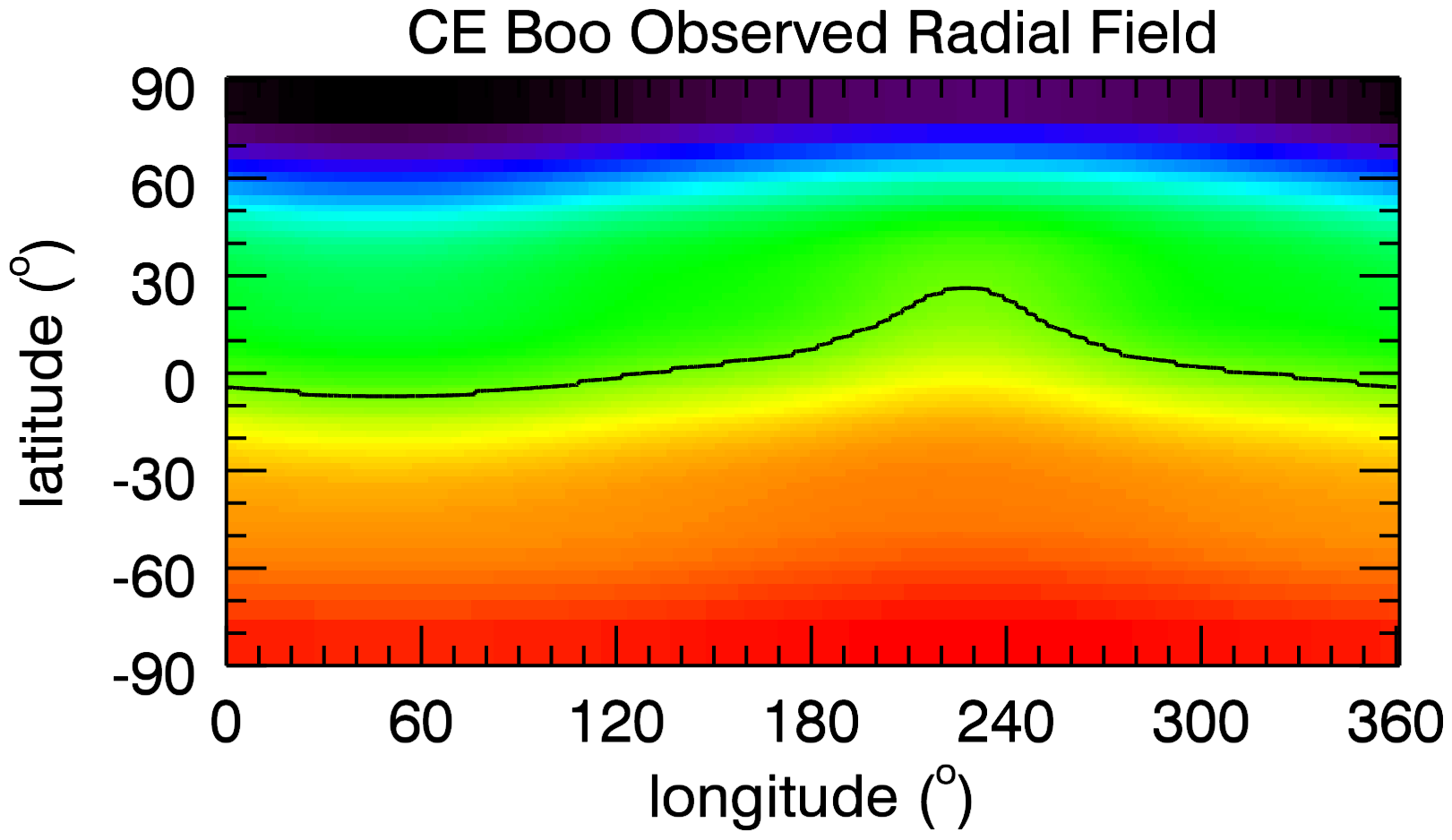}
			\includegraphics[width=57mm]{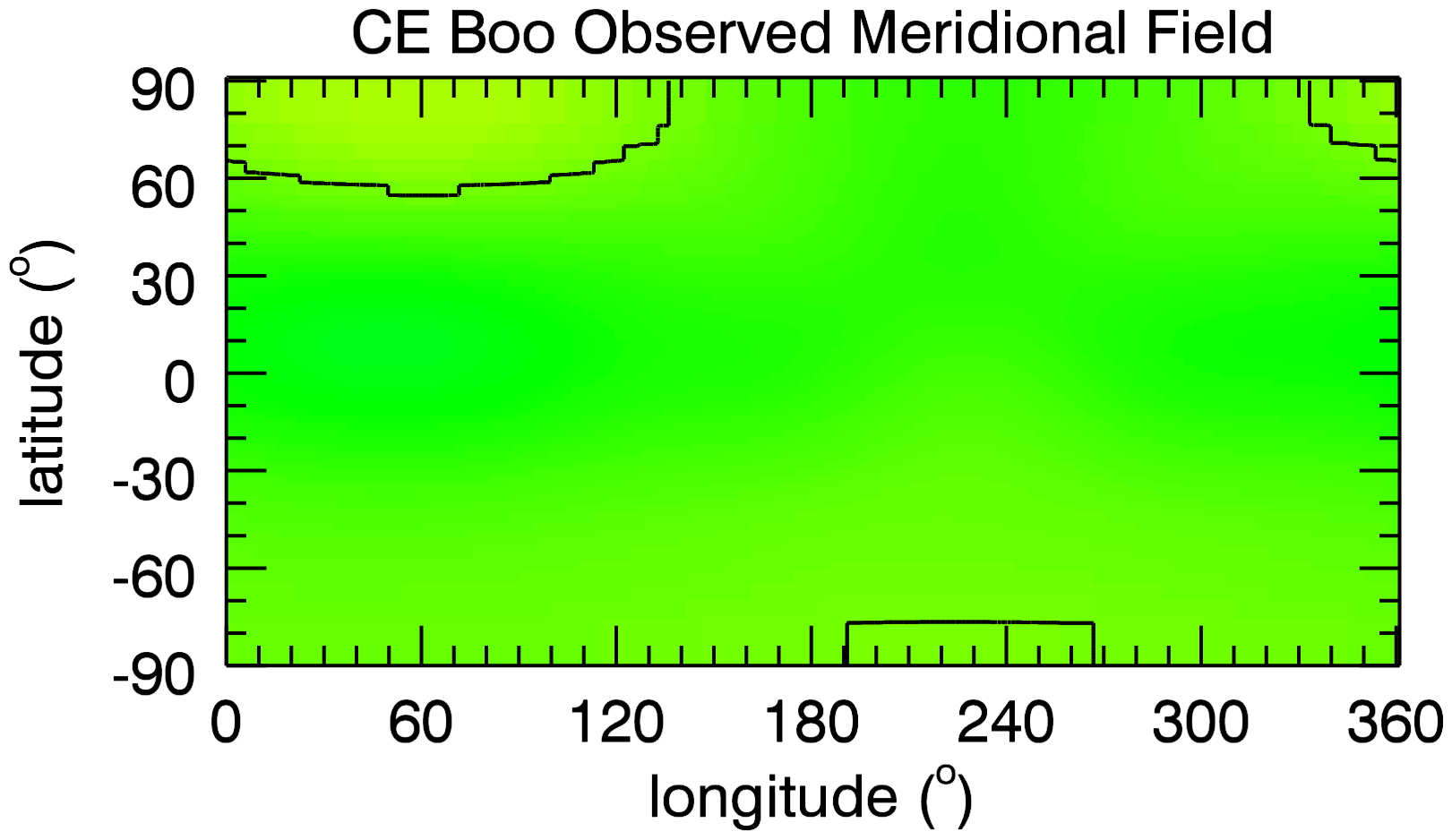}
					\includegraphics[width=57mm]{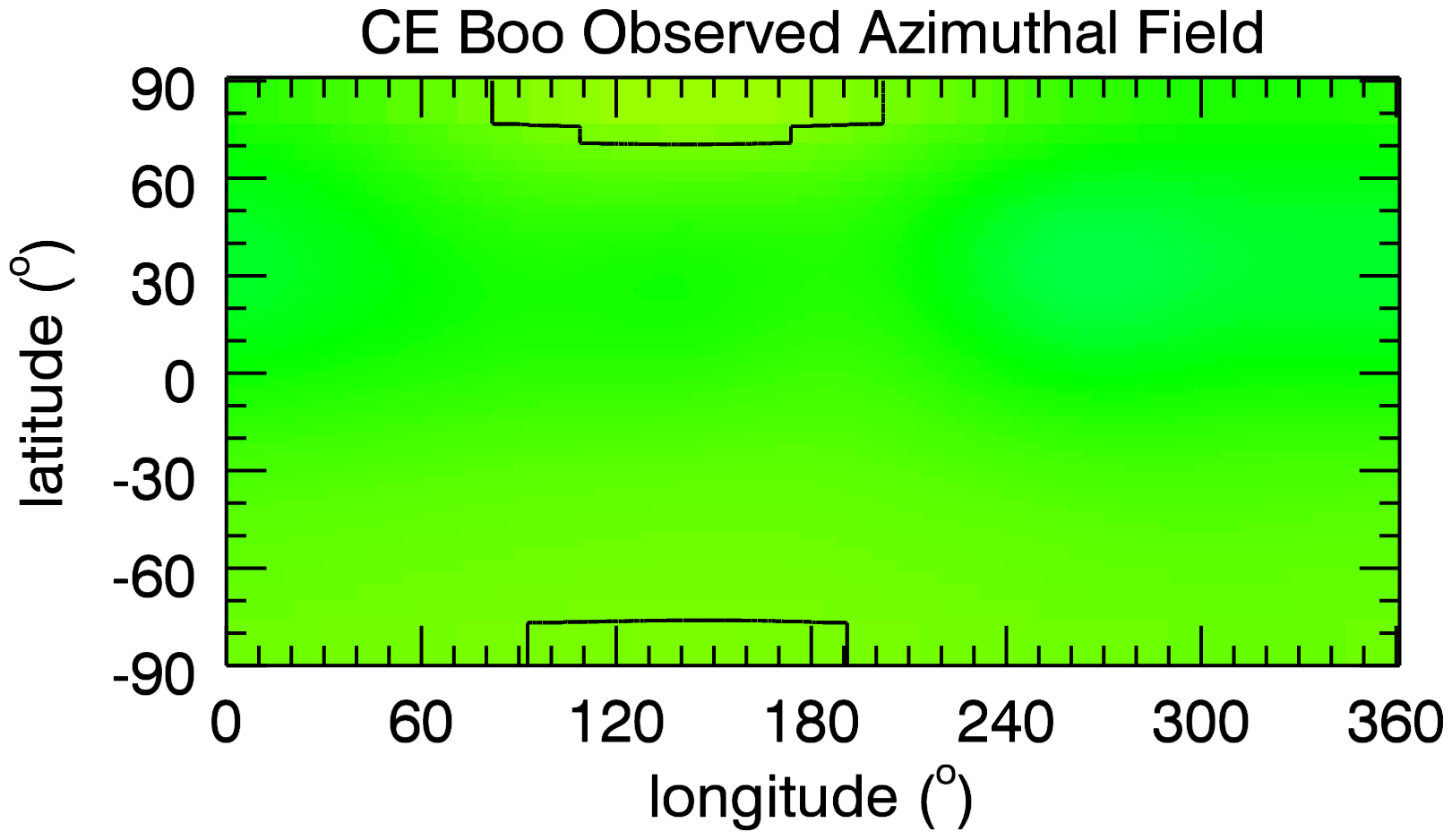}
	    	\includegraphics[width=57mm]{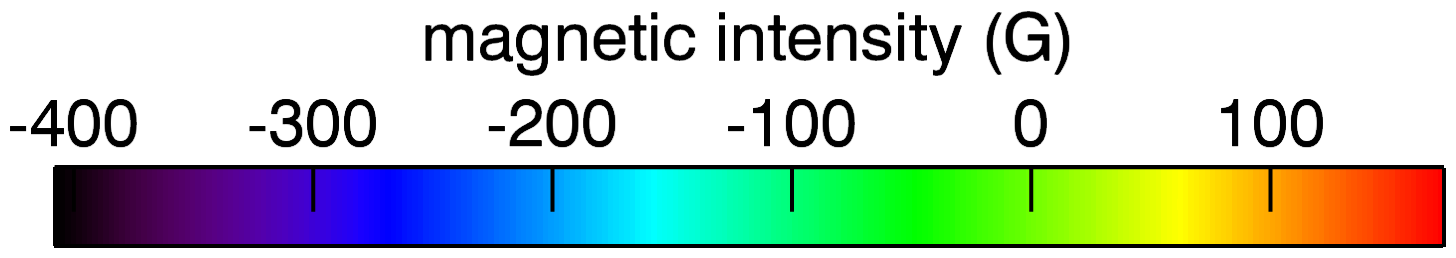}
\caption{Surface magnetic field maps of CE Boo derived from spectropolarimetric observations \citep{morin_earlyM_08}.  The single black line shows the zero-field contour that separates regions of opposite polarity.}
\label{observations_CEBoo}
\end{figure*}


\begin{figure*}
			\includegraphics[width=57mm]{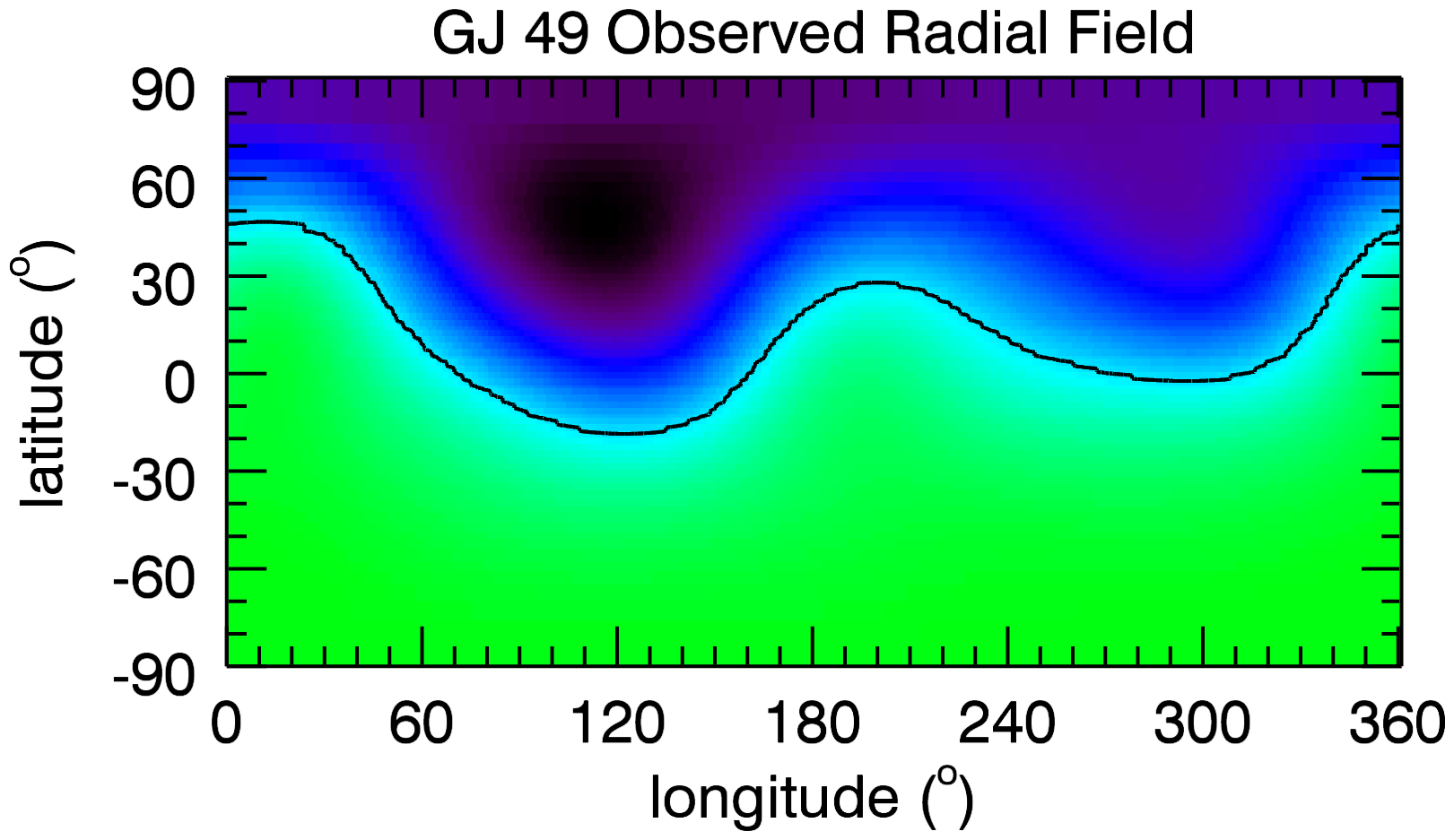}
			\includegraphics[width=57mm]{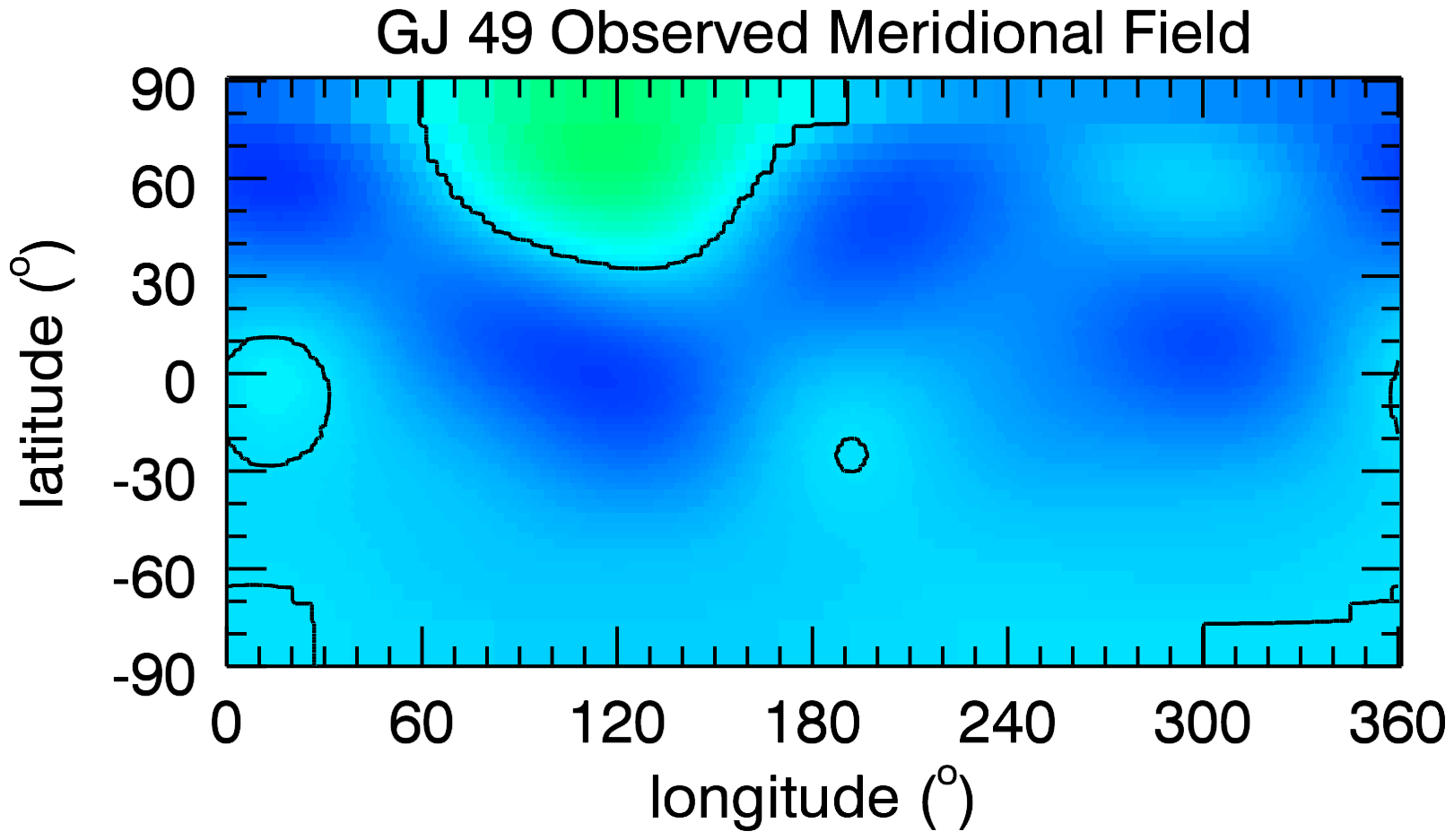}
						\includegraphics[width=57mm]{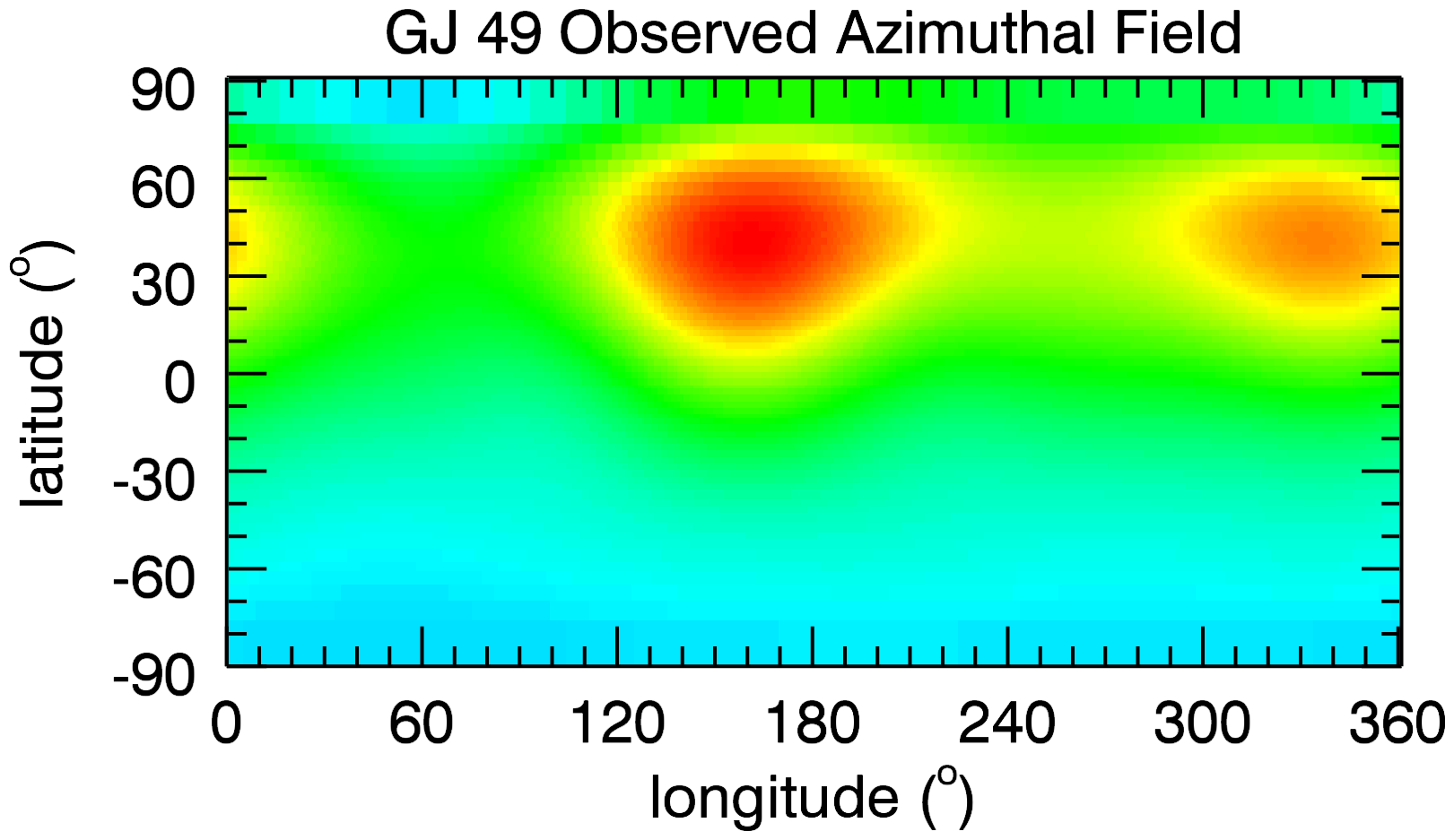}
	    	\includegraphics[width=57mm]{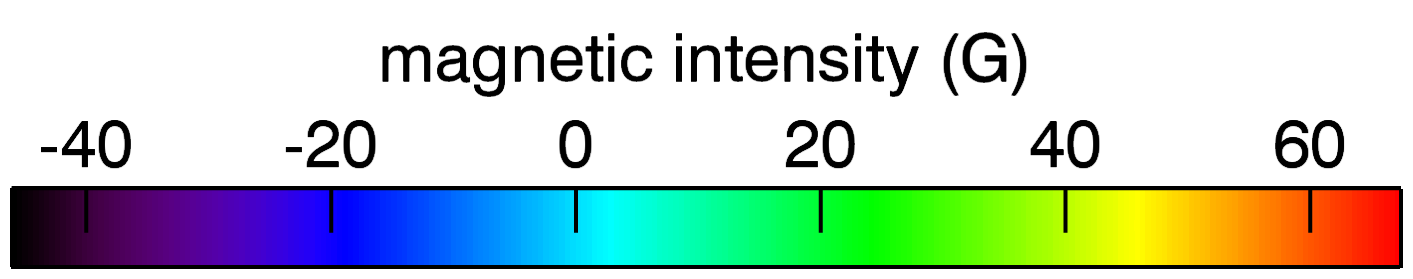}
\caption{Surface magnetic field maps of GJ 49 derived from spectropolarimetric observations \citep{morin_earlyM_08}.  The single black line shows the zero-field contour that separates regions of opposite polarity.}
\label{observations_GJ49}
\end{figure*}


Both of these effects are likely to weaken as the star ages and spins down, generating less magnetic flux and hence producing a weaker wind and reduced X-ray emission \citep{guedel_review_04}. Recent maps of the surface magnetic fields of stars with a range of masses and rotation rates, however, suggest that it is not only the strength of the magnetic field that changes with rotation rate, but also its geometry \citep{morin_earlyM_08,morin_midM_08,petit_toroidal_08,morin_lateM_10}. In contrast to the Sun which shows spots in well-defined ``active latitudes'', solar mass stars that are still in the rapidly-rotating stage typically show very non-solar magnetic fields, with spots that extend over the whole surface, often resulting in a dark polar cap \citep{strassmeier_review_09}. Mixed polarity magnetic flux is seen at all latitudes on these stars. 

Typically these rapidly-rotating stars have X-ray luminosities that are three orders of magnitude greater than that of the Sun, but the extent of the corona that produces this emission is currently unknown. X-ray spectra suggest that their coronae are dense and compact \citep{dupree93, schrijver95,brickhouse98,maggio2000,gudel01,sanz_forcada_abdor_03}. In contrast, the presence of multiple large cool prominences trapped in co-rotation at distances of several stellar radii suggest that their closed magnetic fields, if not their X-ray bright coronae, must be very extended \citep{cameron89eject,cameron89cloud,cameron92alpper,jeffries93,byrne96hkaqr,eibe98re1816,barnes20PZTel,donati20RXJ}. One possible explanation is that the prominences form not within the X-ray bright corona, but in the cusps of helmet streamers that extend out into the stellar wind \citep{jardine05proms}. These prominences typically form in a timescale of 1 day and some 1-10 appear in the observable hemisphere at any time. Their ejection in the stellar equivalent of solar coronal mass ejections not only contributes to the angular momentum loss from the star \citep{aarnio_solarCMEs_11,aarnio_cs16_11} but it will also temporarily enhance the ram pressure of the stellar wind and hence the degree of compression of any planetary magnetospheres.

The surface fields of these young stars show one other very non-solar feature and that is the presence of a strong (sometimes dominant) non-potential component \citep{petit_toroidal_08}. Stellar winds can of course produce azimuthal fields as the escaping wind extracts angular momentum from the star via magnetic torques, but for slow rotators it is unlikely  that this could generate such strong fields at the photospheric level. Several other mechanisms have been proposed to explain the surface azimuthal fields, including the underlying dynamo \citep{donati97abdor95}, and the effect of differential rotation in the presence of a unipolar cap \citep{pointer01evol}. 

For solar mass stars, the surface differential rotation is similar to that of the Sun, but for higher-mass stars the differential rotation can be extreme, with equator to pole lap times as short as 16 days \citep{barnes_DR_05,marsden_DR_05,marsden_DR_06,jeffers_HD171488_11}. The effect that this enhanced shear might have on the coronal and wind dynamics and the possible rate of coronal mass ejections is unknown. For low mass stars the differential rotation is typically weak \citep{morin_midM_08}. The high flaring rate of these stars however suggests that some dynamic process is stressing the coronal field - even although in many cases the large-scale field that is detected with Zeeman-Doppler imaging is close to its potential, or lowest-energy state. 

Some insight into these stellar fields can be gained by considering the changes in the solar magnetic field over the Sun's magnetic cycle. At minimum, the solar field is closest to an aligned dipole, with fast wind streams emerging from the open field regions at the pole and the slow streams emerging from above the low-latitude active regions. As the cycle progresses, more bipoles emerge, contributing to the azimuthal field. These are acted on by diffusion and differential rotation, and their transport towards the poles by the meridional flow eventually reverses the polar polarity. In addition, their net contribution to the azimuthal field causes the axis of the large-scale dipole to move down into the equatorial plane and eventually reverse. This growth of active regions (and associated coronal mass ejections) through the cycle is also accompanied by the extension of the polar coronal holes down towards the equatorial plane. As a result, fast wind streams originate at a range of latitudes and may interact with the slow wind streams to produce ``co-rotating interaction regions'' in the solar wind. These shocks provide a local density enhancement that, combined with the increased number of coronal mass ejections, can modulate the cosmic ray flux at Earth \citep{wang_nonaxisym_11}. Recent models of the variation of the solar wind through its cycle \citep{pinto_solarcycle_11} show that the magnetic torques exerted on the Sun vary significantly through its cycle, giving two orders of magnitude variation in the spin-down time.

By analogy with the Sun, the very active young stars that show predominantly non-axisymmetric and non-potential surface fields may have winds that show a mixture of fast and slow wind streams with coronal mass ejections emerging from a range of latitudes. Indeed, the fact that these stars typically show mixed-polarity flux at all latitudes may suggest that their winds (while showing some characteristics of the solar wind at maximum) are much more extreme than the solar wind.

Most stellar wind models are, however, based on the solar analogy. The simplest early models, such as the traditional Weber-Davies model \citep{weber67},  assumed a split monopole, but more recent work usually initiates MHD simulations from an initial state defined by a ``potential field source surface" model \citep{altschuler69,schatten69}. This approach assumes that the field is potential (i.e. in its lowest energy state) and that at some height above the surface the field lines are opened up by the pressure of the hot coronal gas. This method uses only the radial field component at the surface, neglecting the azimuthal and meridional components. Its advantage is that it is computationally cheap and it provides a unique solution for the magnetic structure.  A recent comparison of the global structure predicted by both the potential field source surface method and the full MHD simulation suggests that the former captures the large-scale structure of the solar coronal field fairly reliably \citep{riley_PFSS_06}.

This approach would not however capture the non-potential nature of the magnetic fields observed at the surfaces of other stars. The purpose of this paper is to explore the effect of this non-potential field on the large-scale structure of the corona and winds of solar-type stars.

\begin{table}
\caption{Stellar and magnetic parameters for CE Boo and GJ 49, taken from \citet{morin_earlyM_08}. The table lists sequentially the stellar name, spectral type, mass, radius, inclination of the rotation axis and the rotation period, and then the field properties: the reconstructed magnetic flux density, and the fractional energy in the poloidal (potential) field.}
\begin{center}
\begin{tabular}{|c|c|c|c|c|c|c|c|c|c|c|c|c|}
\hline
	Star	&	Sp. Type		&	M$_\star$		&	R$_\star$	&	i				&	P$_{rot}$		&					$<$B$>$	&	Pol	\\
			&	 		&	(M$_\odot$)		&	(R$_\odot$)	&	($^\circ$)		&	(d)				&	(G)			&			&\\
\hline
CE Boo	&	M2.5	&	0.43				&		0.48		&	45				&	14.7			&	103		&	0.95	\\
	GJ 49	&	M1.5	&	0.51				&		0.57		&	45				&	18.6			&	27			&	0.48	\\
\hline
\end{tabular}
\end{center}
\label{parameters}
\end{table}%

\section{The surface magnetograms}
In order to study the effects of the non-potential field on the coronal structure and dynamics we choose to compare two stars (CE Boo and GJ 49) that are similar in rotation rate but with slightly different masses. One has a surface field that is close to potential, while the other has a significant non-potential component. Both stars are slow rotators, so rotational effects are minimal. In addition, the inclination of the rotation axes of both stars to the line of sight is the same, so the magnetic fields of both stars are seen in the same orientation. The stellar parameters are shown in Table \ref{parameters}. 

We choose initially to compare the static coronal structures that are found by assuming either that the field is purely potential, or that it has both potential and non-potential components. These extrapolations can be used as the initial condition for a full MHD solution. Since we are particularly interested in the non-potential field, we also explore the possibility that for the more rapidly-rotating stars it is the rotational stressing of the surface field by the action of the wind that causes the field to depart from a potential state. We therefore perform one simulation of the wind of GJ 49 with an artificially decreased stellar rotation period of 0.6 days and compare this to the wind parameters found with the observed rotation period of 18.6 days.

The input for the static extrapolation is taken from \citet{morin_earlyM_08}. The surface magnetic field of both stars were modelled with Zeeman-Doppler imaging from time-series of spectropolarimetric observations collected over approximately 2 consecutive stellar rotations. For spatially unresolved sources, due to the mutual cancellation of contributions from neighbouring
regions of opposite polarities to the polarized signal, spectropolarimetric measurements can only probe the large-scale component of magnetic fields (see e.g. \citet{morin12}). The maximum degree $\ell$ of modes that can be reconstructed with ZDI depends on the star's projected rotational velocity.
For slow rotators such as CE Boo and GJ 49, the reconstruction is limited to modes with order $\ell \leq 8$. As there is no unique solution to the ZDI problem, a regularization scheme has to be used. A maximum entropy solution corresponding to the lowest magnetic energy content is used. It is
optimal in the sense that any feature present in the map is actually required to fit the data. Although this method does not allow us to derive formal error bars on the reconstructed maps, numerical experiments have shown that ZDI is a robust method \citep{donati97,morin_lateM_10}. 

This reconstructed field is expressed as a sum of a poloidal and toroidal field \citep{mestel_book_99}. The poloidal component captures the potential contribution to the total field, that is the component that is in its lowest-energy state. The toroidal component lies on the surfaces of concentric spheres and captures the non-potential component of the total field. It is this component that is associated with the electric currents in the corona and which describe the free energy that is available to power, for example, stellar flares and coronal mass ejections. These two components of the surface field can be expressed as linear combinations of spherical harmonics \citep{donati06tausco}. Thus the radial, meridional and azimuthal field components at the stellar surface can be written in spherical co-ordinates $(r,\theta,\phi)$ as\footnote{We note that in \citet{donati06tausco}, the radial field is positive outwards, the azimuthal field is positive in the direction of stellar rotation (i.e. increasing longitude or decreasing rotation phase) and the meridional field is positive when pointing to the visible pole.}

\begin{figure*}
	\includegraphics[width=88mm]{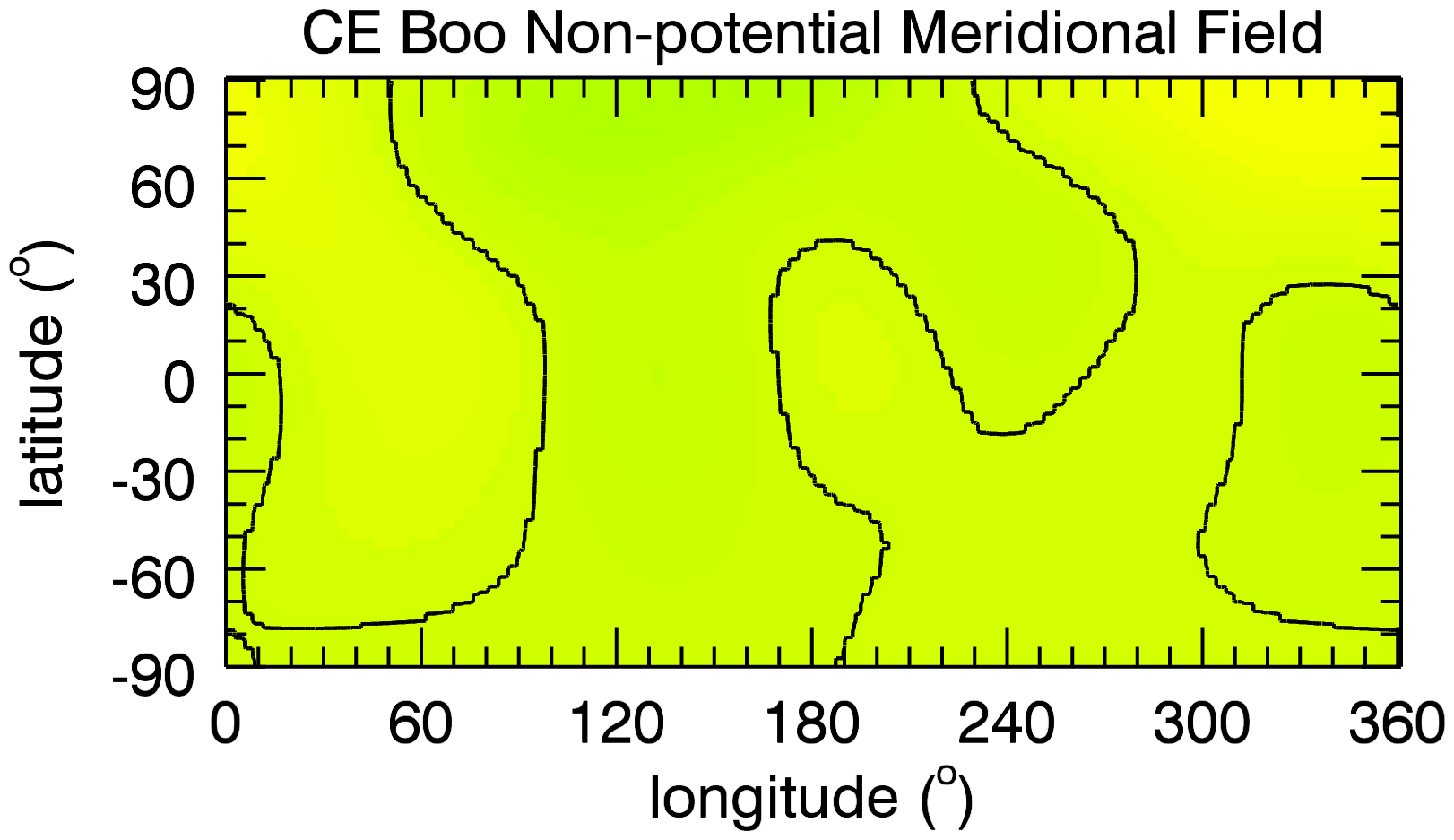}
	\includegraphics[width=88mm]{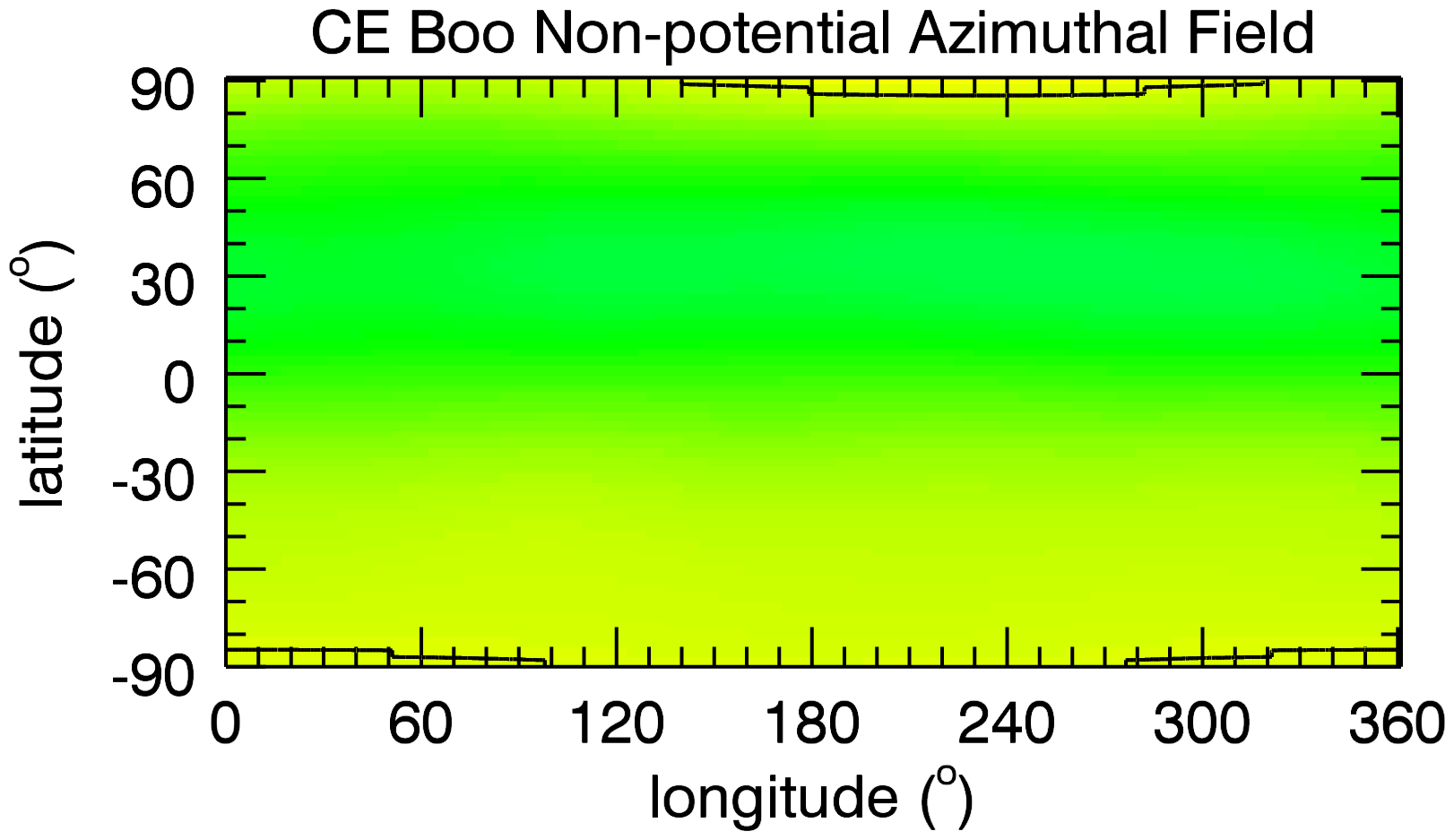}
	\includegraphics[width=88mm]{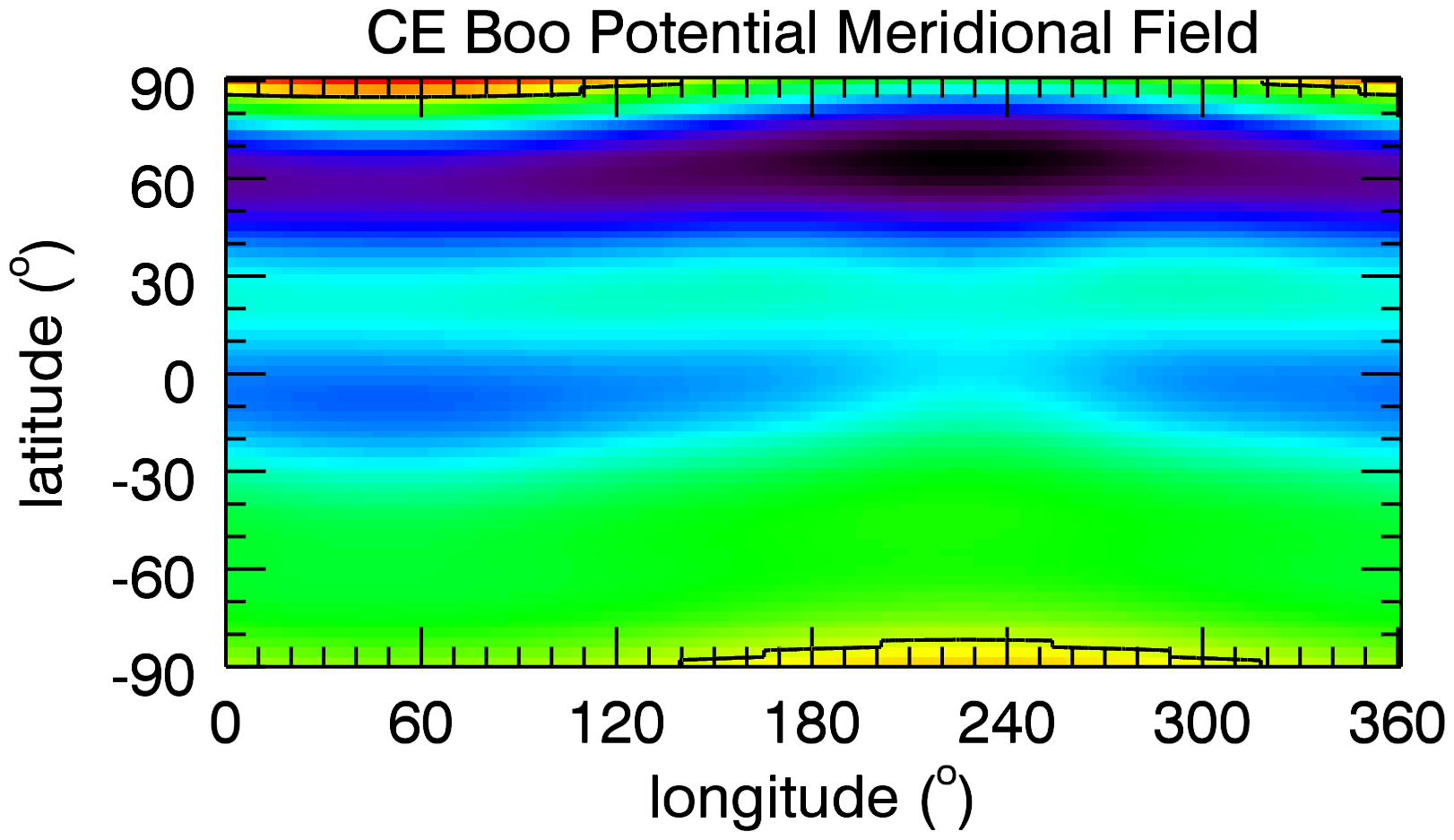}
	\includegraphics[width=88mm]{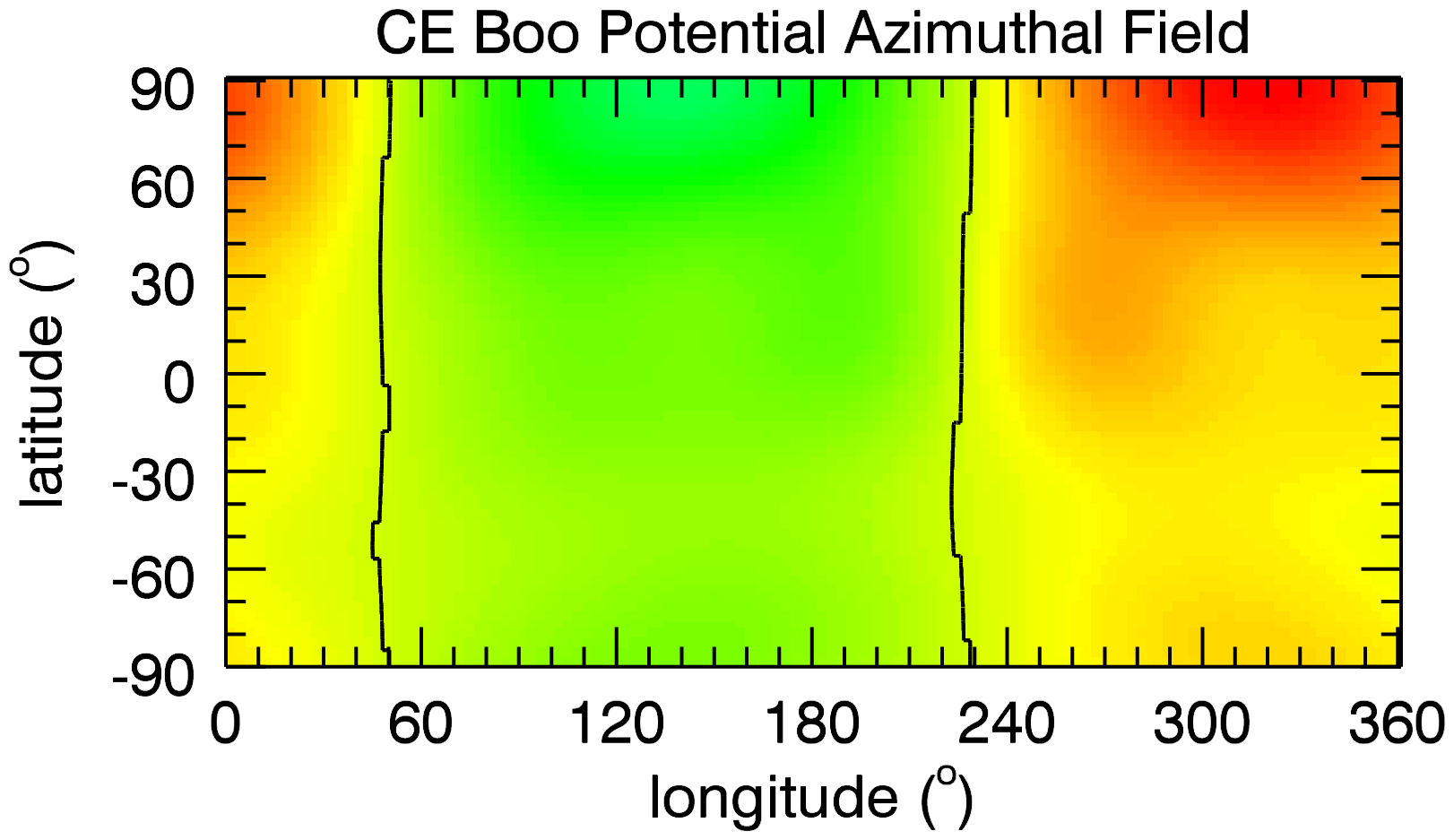}
	    	\includegraphics[width=88mm]{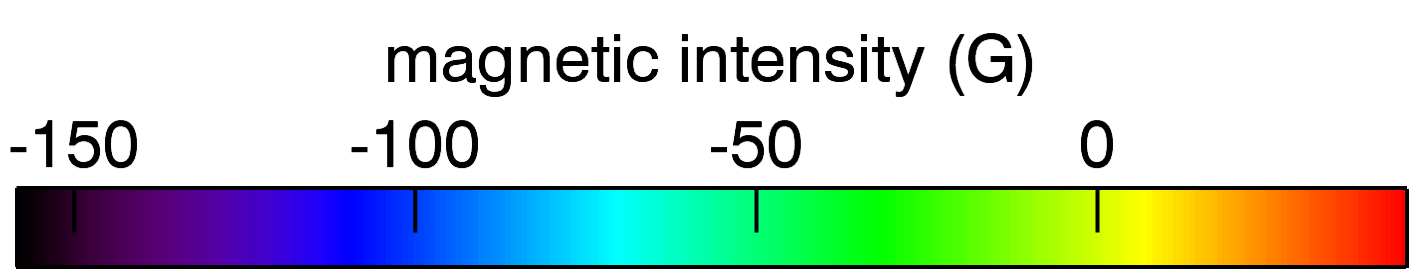}
\caption{The static solution for the surface magnetic field of CE Boo, divided into its different components. 
The meridional component is shown in the left column and the azimuthal component in the right column.  The top row shows the non-potential contribution and the bottom row the potential contribution to the total field. 
The single black line shows the zero-field contour which therefore separates regions of opposite polarity.}
\label{CEBoo_static}
\end{figure*}


\begin{figure*}
%
	\includegraphics[width=88mm]{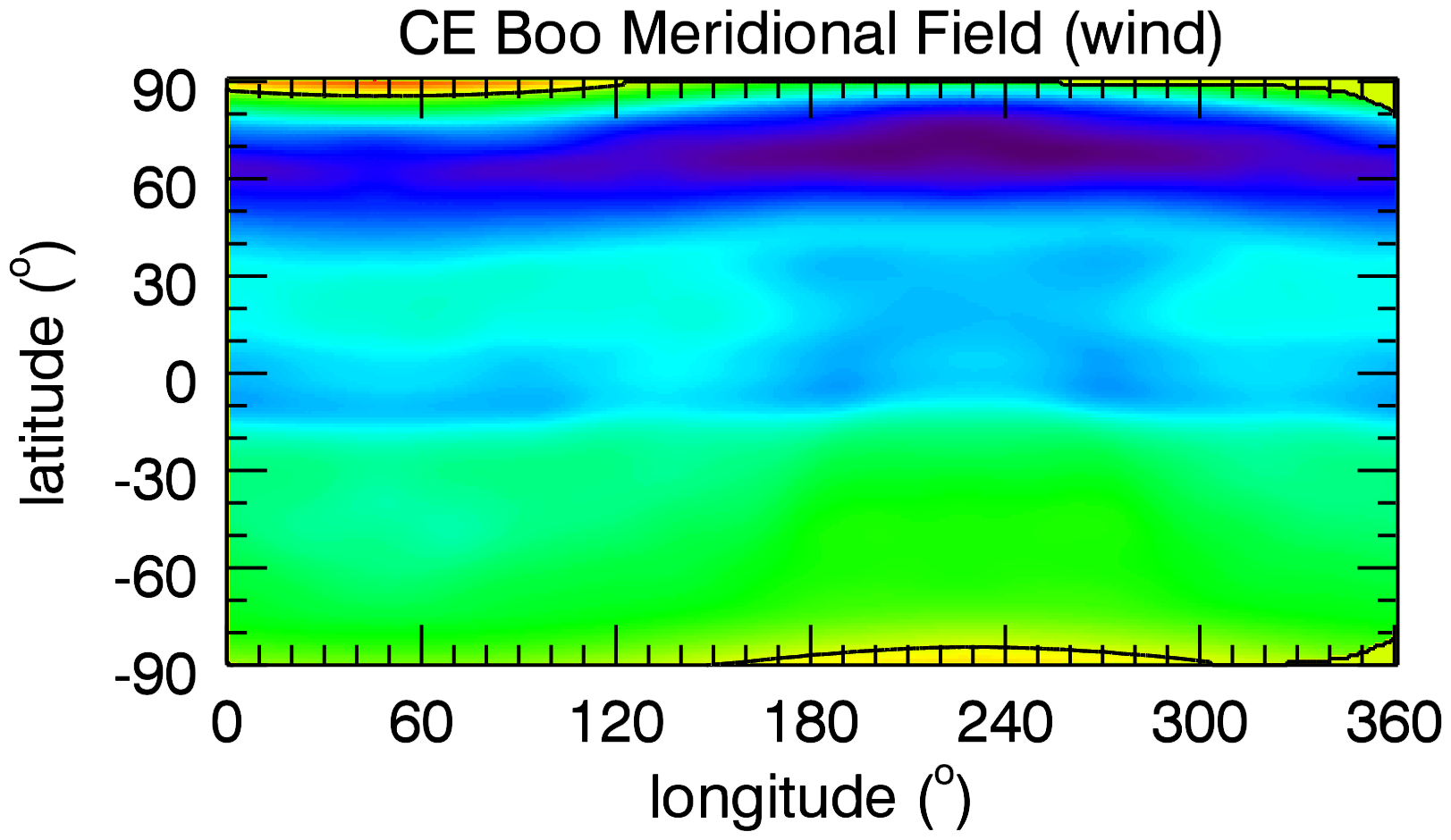}
	 \includegraphics[width=88mm]{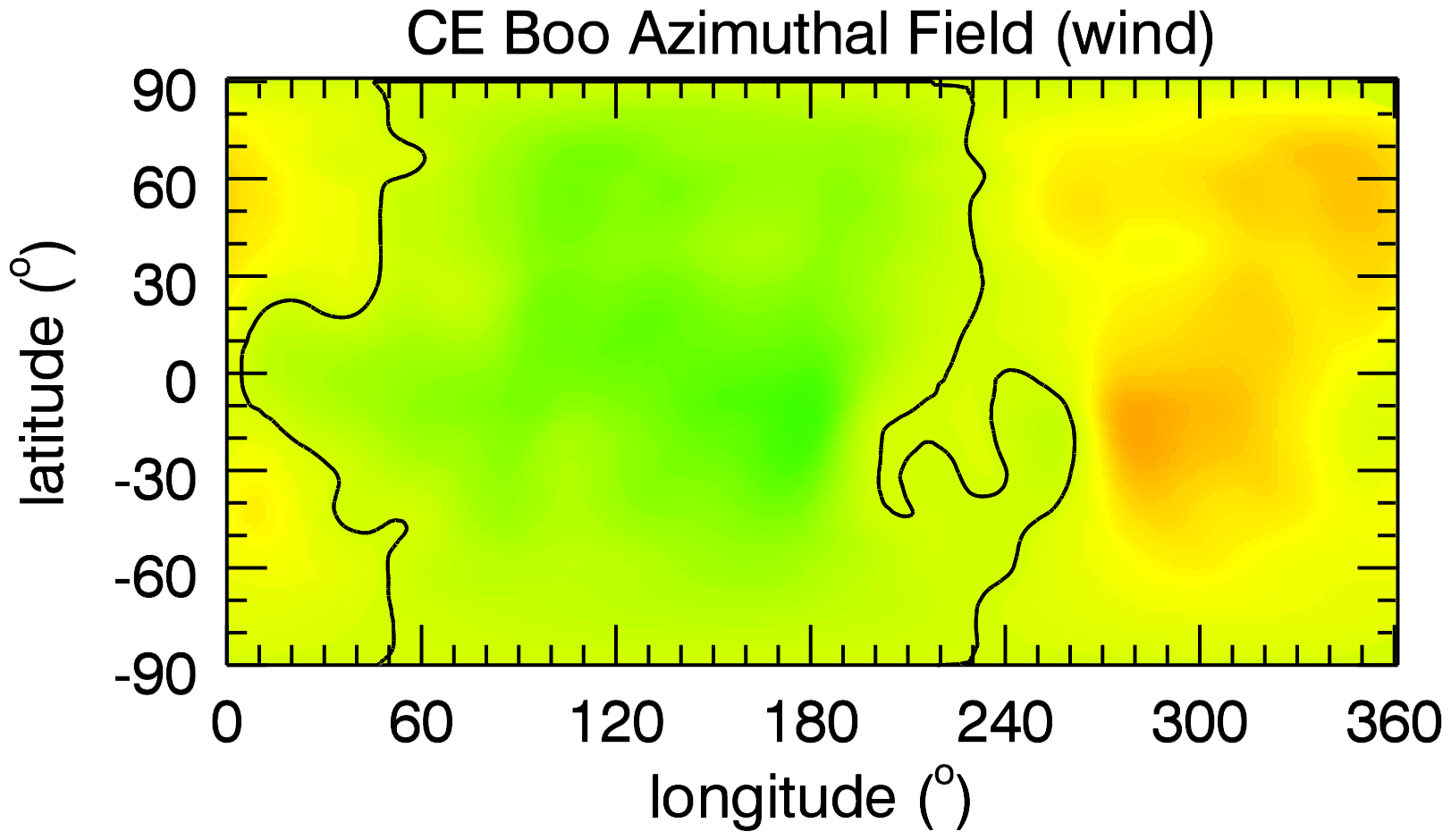}
	    	\includegraphics[width=88mm]{pix_for_paper/Figs2_3_cbar.png}
\caption{The wind solution for the surface magnetic field of CE Boo, divided into its meridional (left column) and azimuthal (right column) components.  The single black line shows the zero-field contour which therefore separates regions of opposite polarity.}
\label{CEBoo_wind}
\end{figure*}


\begin{equation}
B_r =  -\sum^N_{l=1}\sum^l_{m=-l} 
                    \alpha_{lm}c_{lm}P_{lm}(\theta)e^{im\phi} 
\label{br_obs}
\end{equation}
\begin{equation}
B_\theta =  -  \sum^N_{l=1}\sum^l_{m=-l} 
    \left[ 
          \beta_{lm}\frac{c_{lm}}{(l+1)}\frac{dP_{lm}(\theta)}{d\theta} 
    +  \gamma_{lm} \frac{c_{lm}}{(l+1)} \frac{P_{lm}(\theta)}{\sin\theta}im    
    \right] e^{im\phi}
\label{btheta_obs} 
\end{equation}
\begin{equation}
B_\phi =  - \sum^N_{l=1}\sum^l_{m=-l} 
                    \left[ 
                         \beta_{lm}\frac{c_{lm}}{(l+1)}\frac{P_{lm}(\theta)}{\sin\theta} im  
                         - \gamma_{lm}  \frac{c_{lm}}{(l+1)}\frac{dP_{lm}(\theta)}{d\theta}    
                    \right] e^{im\phi}   
\label{bphi_obs}      
\end{equation}
where $l$ and $m$ are the degree and order respectively,
\begin{equation}
c_{lm}=\sqrt{\frac{(2l +1)}{4\pi}\frac{(l - m)!}{(l + m)!}}
\end{equation}
 and $P_{lm}(\theta)$ denotes the associated Legendre functions. The potential terms are those with coefficients $\alpha_{lm}$ or $ \beta_{lm}$, while the non-potential terms are those with coefficients $\gamma_{lm}$. Clearly, then, in the limit $\gamma_{lm} \rightarrow 0$ we recover a purely potential field.

The corresponding surface magnetic maps from \citet{morin_earlyM_08} are reproduced in Figs. \ref{observations_CEBoo} and \ref{observations_GJ49}. In both cases, the radial and meridional fields look very similar to a dipole, but particularly in the case of GJ 49, there is a significant azimuthal field that is unidirectional at low to mid latitudes. This is a clear signature of a non-potential field.  

\section{The static coronal magnetic field}
In order to determine the coronal structure that corresponds to these surface fields we need to make some assumptions about the nature of the coronal field. The simplest assumption is that the field is potential, or in its lowest energy state and is determined simply by the coefficients $\alpha_{lm}$ and $ \beta_{lm}$ in Eqns. (\ref{br_obs} - \ref{bphi_obs}). This is the starting point for many extrapolations of the solar magnetic field. If we wish to determine the distribution of electric currents in the corona, however, we need to allow for the non-potential components that are described by the coefficients $\gamma_{lm}$.

\begin{figure*}
	\includegraphics[width=88mm]{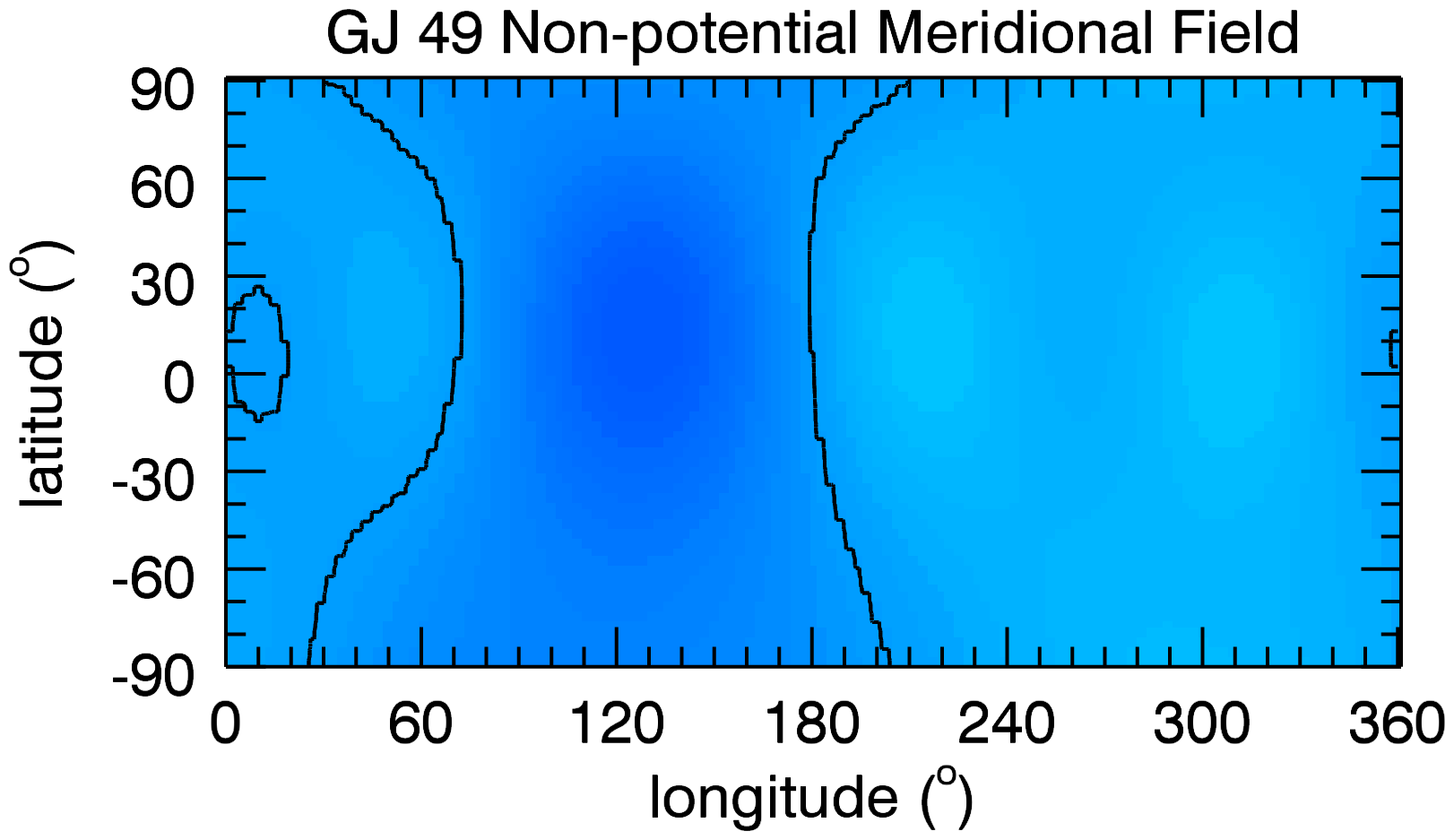}
	\includegraphics[width=88mm]{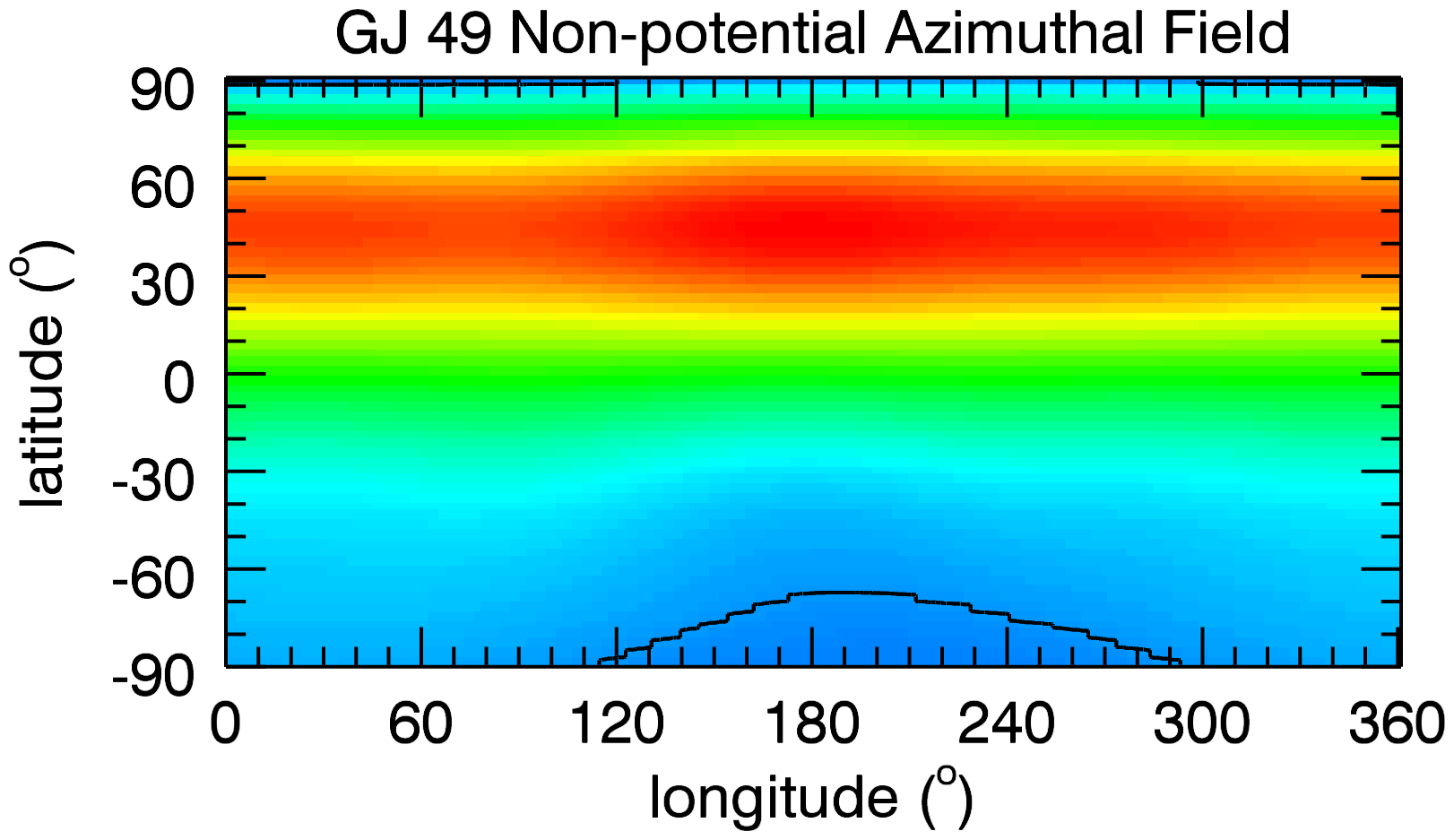}
	\includegraphics[width=88mm]{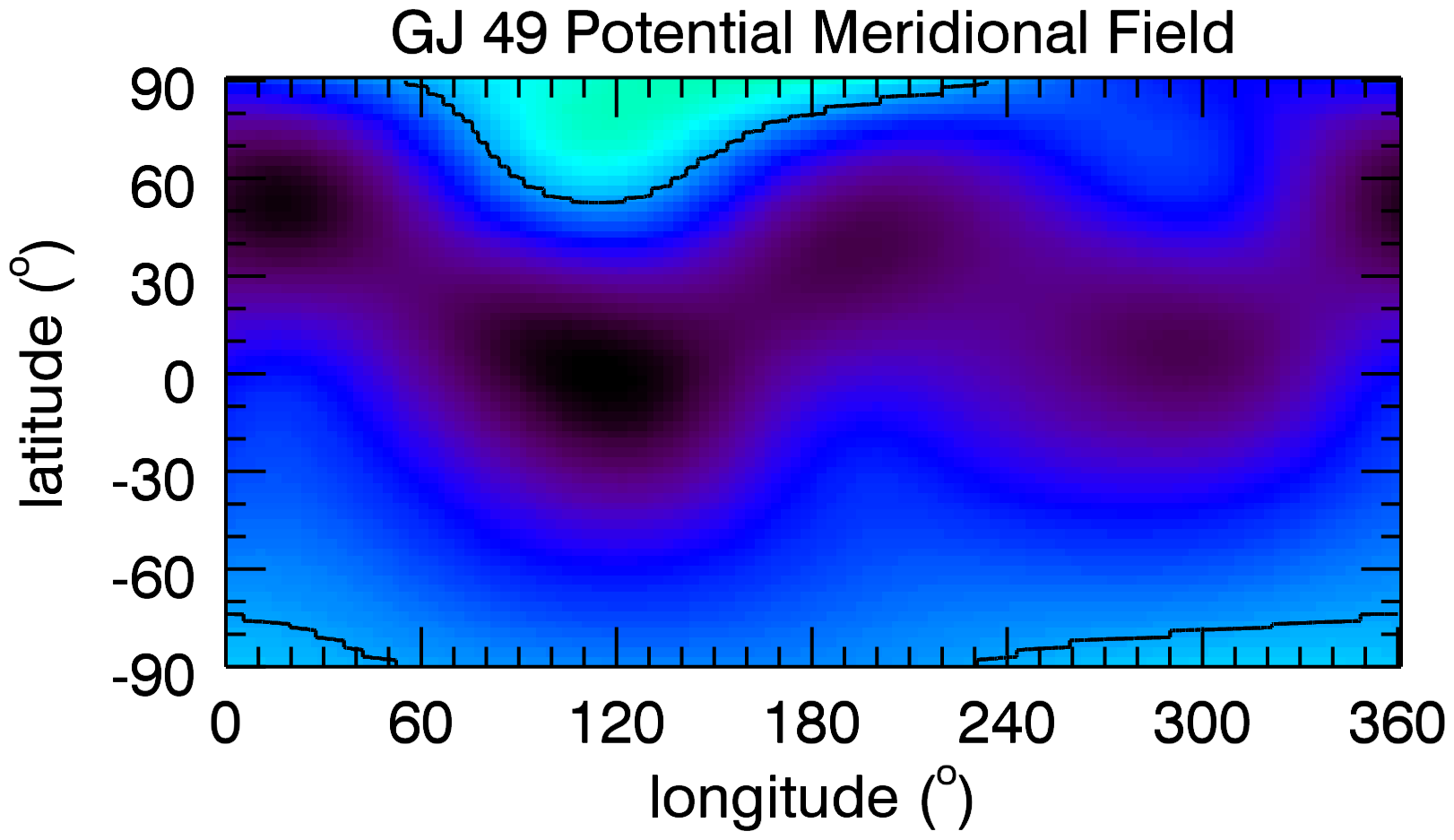}
	\includegraphics[width=88mm]{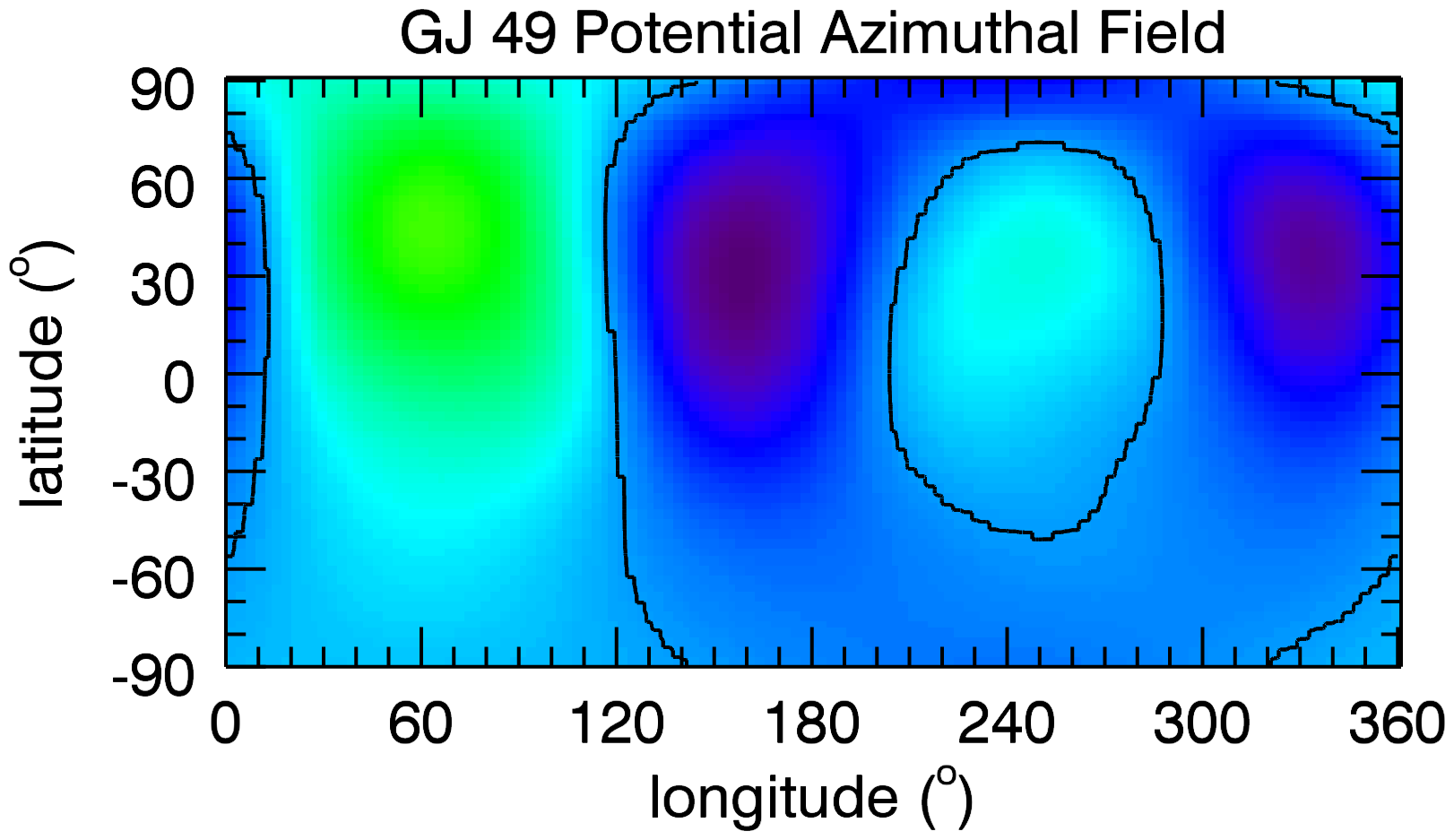}
	    	\includegraphics[width=88mm]{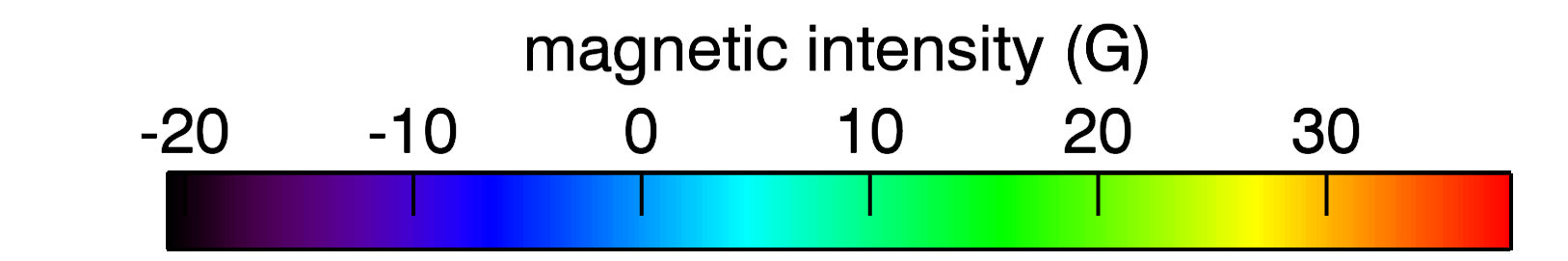}
\caption{The static solution for the surface magnetic field of GJ 49, divided into its different components. The meridional component is shown in the left column and the azimuthal component in the right column.  The top row shows the non-potential contribution and the bottom row the potential contribution to the total field. The single black line shows the zero-field contour which therefore separates regions of opposite polarity.}
\label{GJ49_static}
\end{figure*}


\begin{figure*}
%
			\includegraphics[width=88mm]{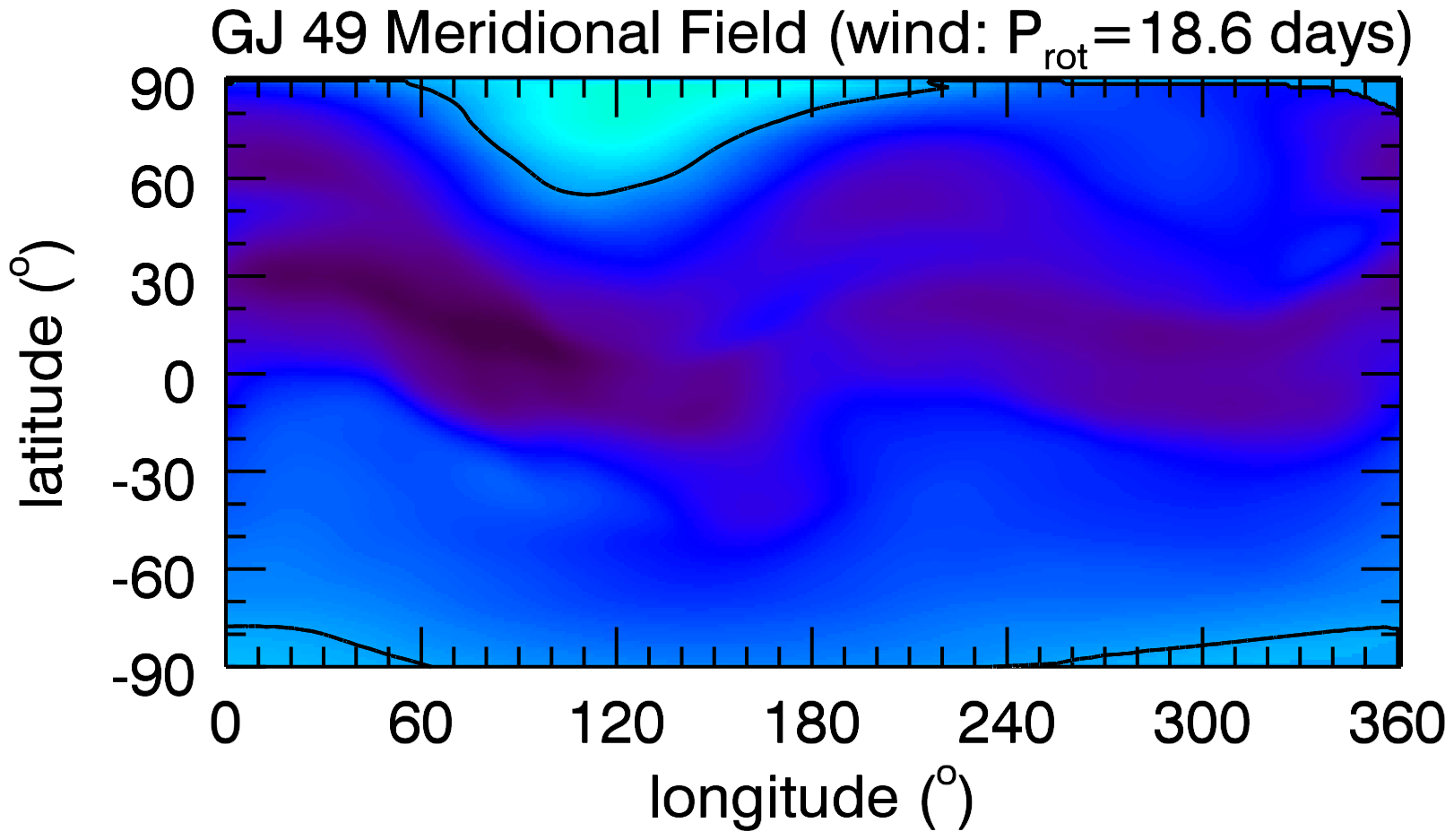}
           	\includegraphics[width=88mm]{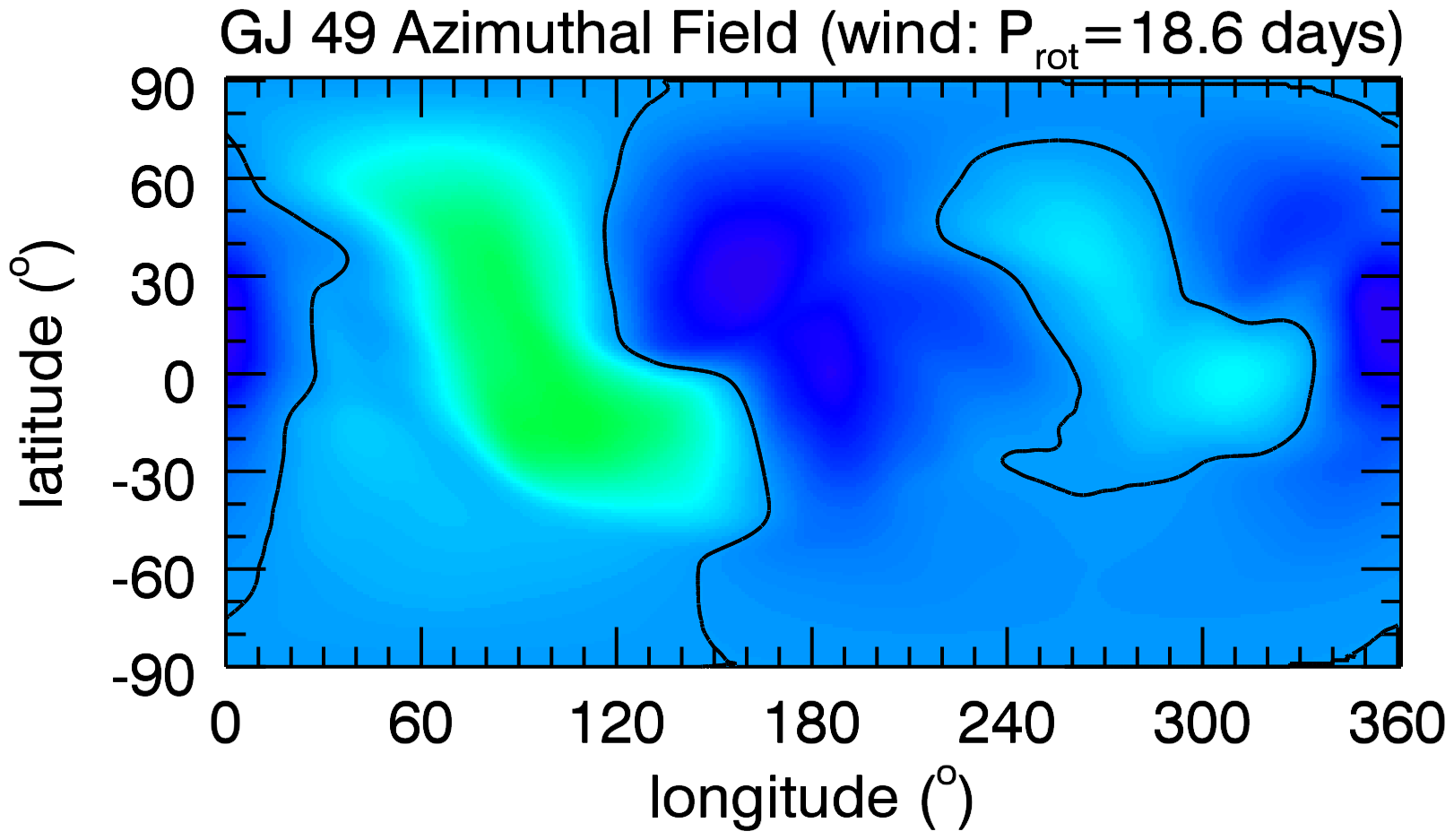}
			\includegraphics[width=88mm]{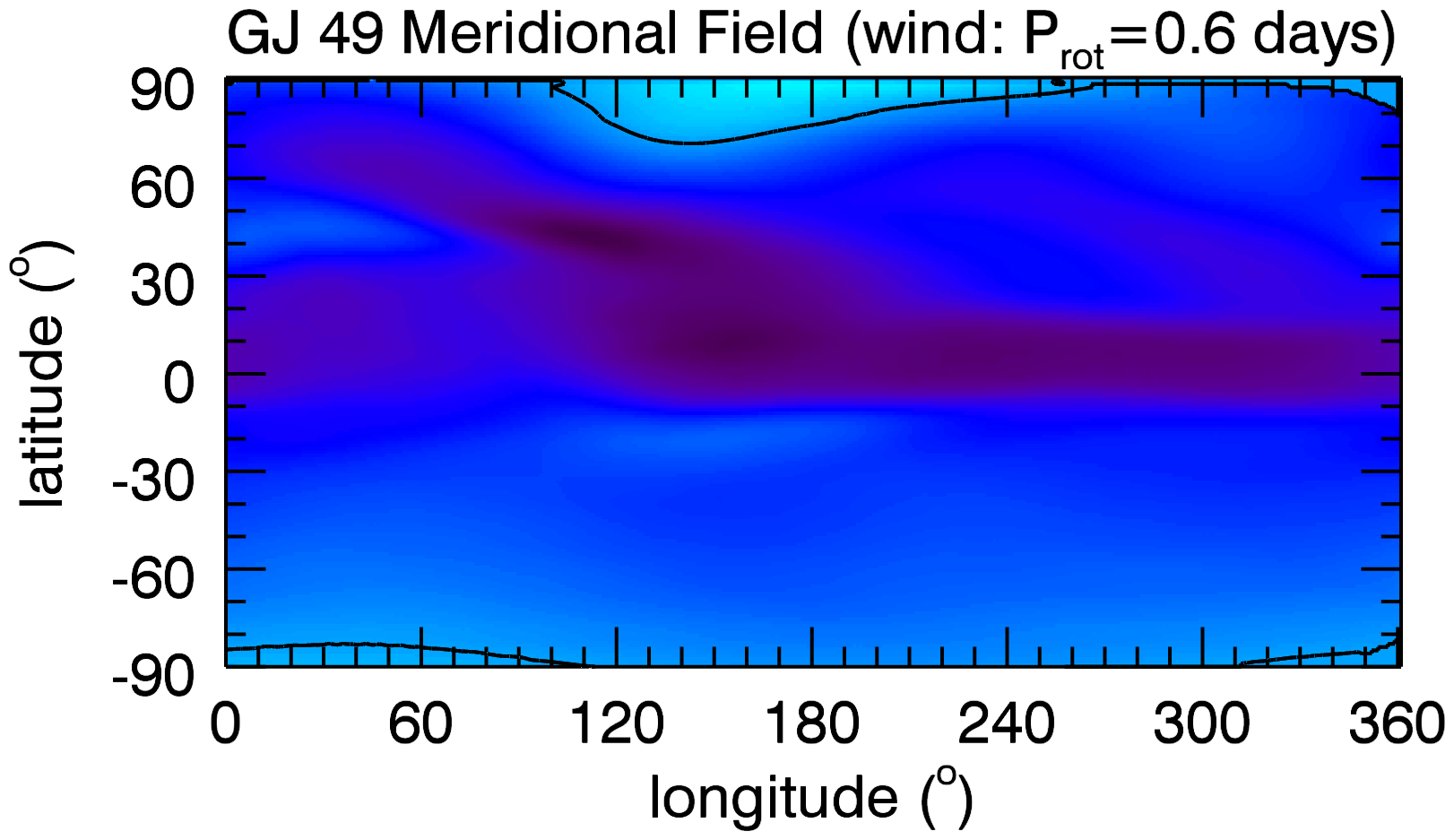}
           	\includegraphics[width=88mm]{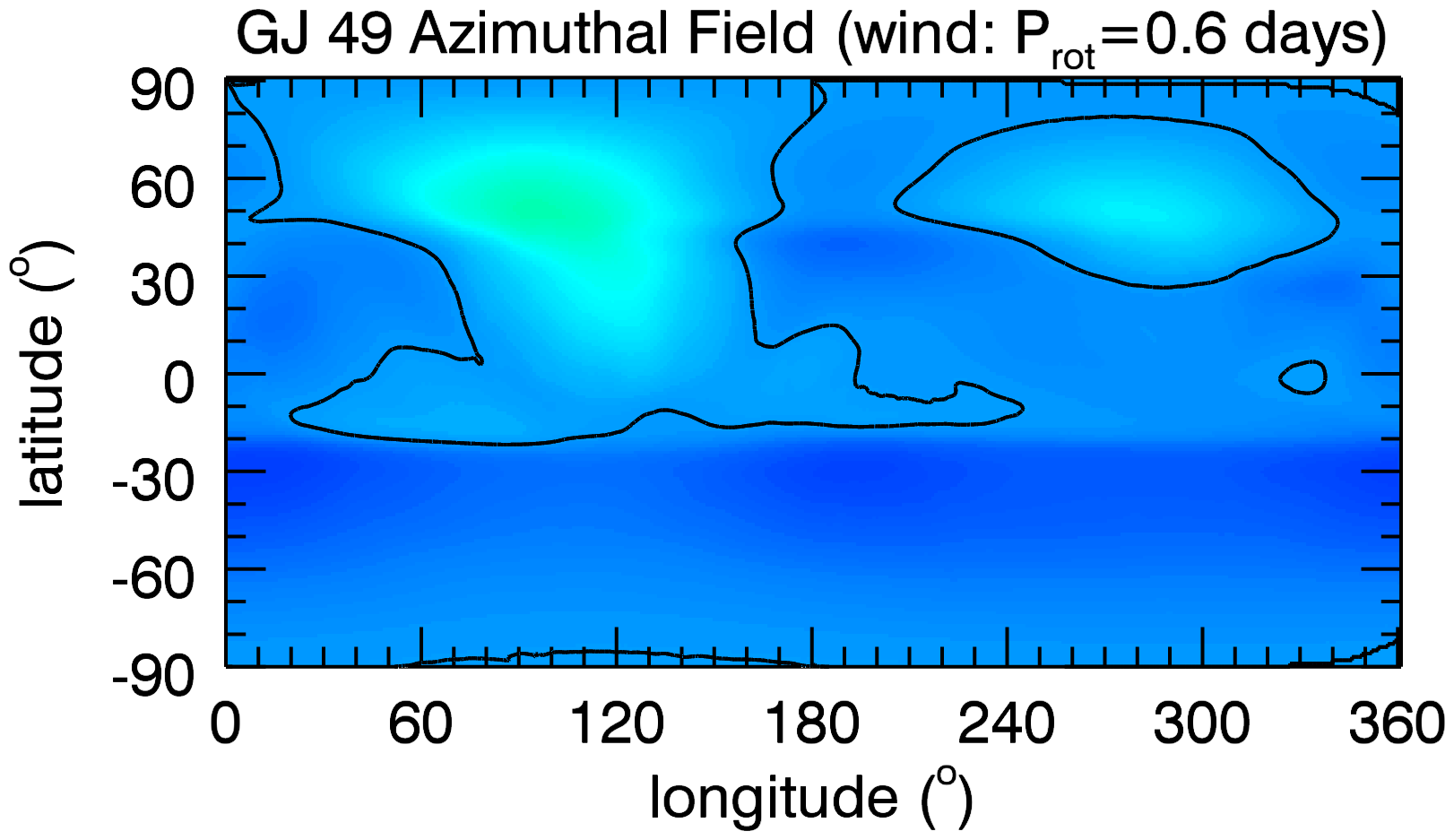}
	    	\includegraphics[width=88mm]{pix_for_paper/Figs4_5_cbar.png}
	
\caption{The wind solution for the surface magnetic field of GJ49, divided into its meridional (left column) and azimuthal (right column) components.  The top row shows the result of assuming the observed stellar rotation period of 18.6 days, while the bottom row shows the result of assuming  a stellar rotation period artificially decreased to 0.6 days. The single black line shows the zero-field contour which therefore separates regions of opposite polarity.}
\label{GJ49_wind}
\end{figure*}


\subsection{Potential field extrapolation}
\label{potential_section}
We begin by calculating the contribution to the total field that is potential. We write  $\underline{B}^{pot}$ in terms of a flux function $\Psi$
such that $\underline{B}^{pot} = -\underline{\nabla} \Psi$ and the condition that the
field is potential ($\underline{\nabla}\times\underline{B}^{pot} =0$) is then satisfied
automatically.  The condition that the field is divergence-free then
reduces to Laplace's equation $\underline{\nabla}^2 \Psi=0$ with solution 
in spherical co-ordinates $(r,\theta,\phi)$
\begin{equation}
 \Psi = \sum_{l=1}^{N}\sum_{m=-l}^{l} [a_{lm}r^l + b_{lm}r^{-(l+1)}]
         P_{lm}(\theta) e^{i m \phi},
\end{equation}
where all radii are scaled to a stellar radius and the associated Legendre functions are once again denoted by $P_{lm}$.  The
two unknowns are therefore the coefficients $a_{lm}$ and $b_{lm}$. One of these can be determined by imposing the
radial field at the surface from the Zeeman-Doppler maps. In order to determine the second unknown, we select a particular form of potential field that has the useful property that at some radius, all the field lines are open. This mimics the effect of the outward pressure of the hot coronal gas pulling open field lines to form the stellar wind. Thus, at some normalised radius $R_s$ above the surface (known as the {\it source surface}) the field becomes radial and hence $B_\theta (R_s) = B_\phi (R_s) = 0$. As a result,
\begin{equation}
b_{lm}=-a_{lm} R_s^{2l+1}
\end{equation}
and we may write
\begin{equation}
B_r ^{pot}=  \sum^N_{l=1}\sum^l_{m=-l} 
                    B_{lm}P_{lm}(\theta)f_l(r,R_s)r^{-(l+2)}e^{im\phi} 
\label{br_pot}
\end{equation}
\begin{equation}
B_\theta^{pot}   =   -   \sum^N_{l=1}\sum^l_{m=-l} 
            B_{lm}\frac{dP_{lm}(\theta)}{d\theta}g_l(r,R_s)r^{-(l+2)}    e^{im\phi}
\label{btheta_pot}
\end{equation}
\begin{equation}
B_\phi^{pot}   =  - \sum^N_{l=1}\sum^l_{m=-l} 
                        B_{lm}\frac{P_{lm}(\theta)}{\sin\theta} im g_l(r,R_s)r^{-(l+2)} e^{im\phi}          
\label{bphi_pot}
\end{equation}
where the functions $ f_l(r,R_s)$ and $g_l(r,R_s)$  which describe the influence of the source surface (and hence the wind) on the magnetic field structure are given by
\begin{equation}
 f_l(r,R_s) = \left[ 
        \frac{l+1+ l(r/R_s)^{2l+1}}{l+1+l(1/R_s)^{2l+1}}
            \right]
\end{equation}
\begin{equation}
 g_l(r,R_s) =  \left[
       \frac{1 - (r/R_s)^{2l+1}}{l+1+l(1/R_s)^{2l+1}}
               \right].
\end{equation}
In the limit where the source surface is large (i.e. the magnetic field is completely closed), we recover the familiar multipolar expansions for a magnetic field. This limit corresponds to $R_s\rightarrow \infty$ and 
\begin{equation}
 f_l(1) \rightarrow 1
\end{equation}
\begin{equation}
 g_l(1) \rightarrow \frac{1}{l+1}.
\end{equation}
The coefficients $B_{lm}$ are determined by the surface radial field that is derived from the Zeeman-Doppler maps (i.e. by the values of $\alpha_{lm}$ in (\ref{br_obs})). This is known as the {\it Potential Field Source Surface} method. It was originally developed for extrapolating the Sun's coronal field from solar magnetograms  \citep{altschuler69}. We use a code originally developed by \citet{vanballegooijen98} (see also \citet{jardine02structure}).

\begin{figure*}
		\includegraphics[width=85mm]{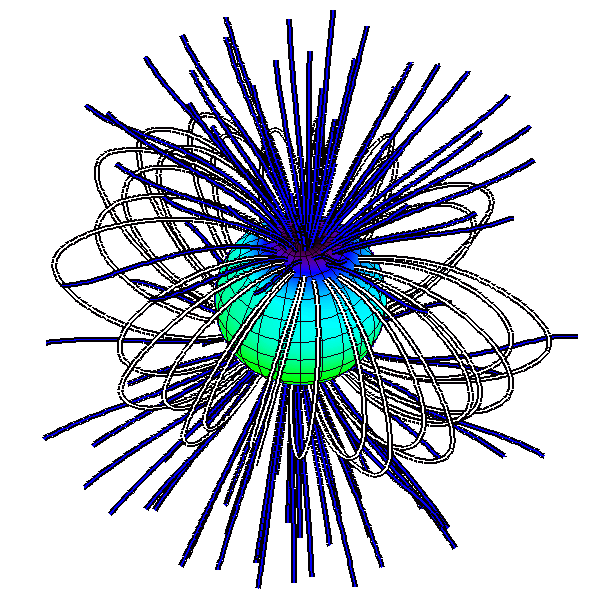}
		\includegraphics[width=85mm]{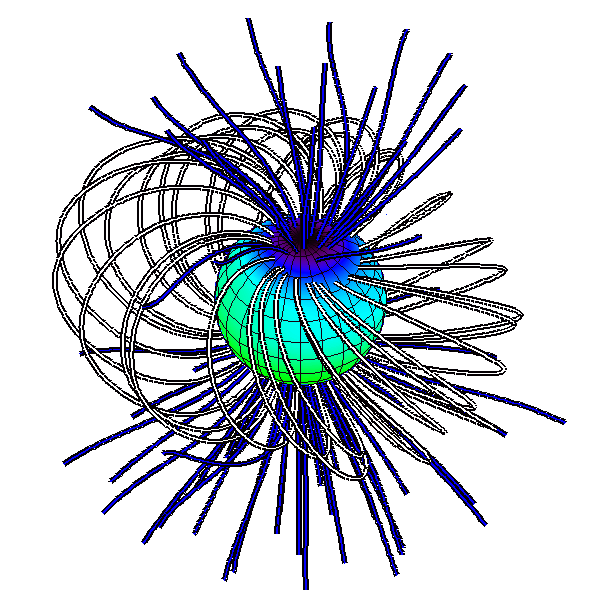}
		\includegraphics[width=85mm]{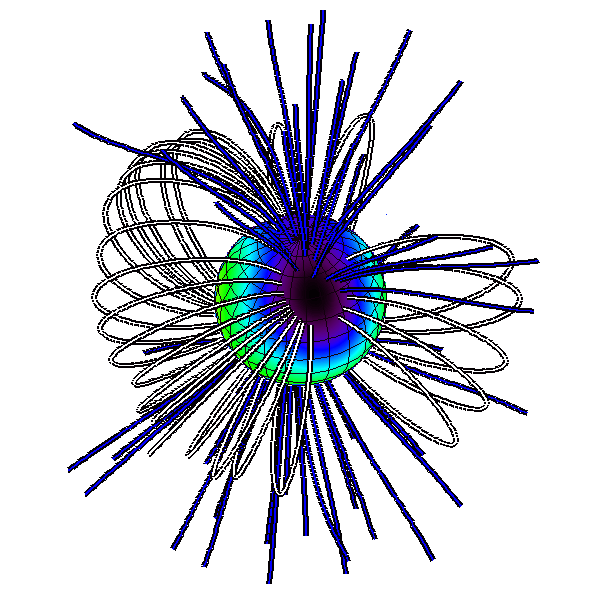}
		\includegraphics[width=85mm]{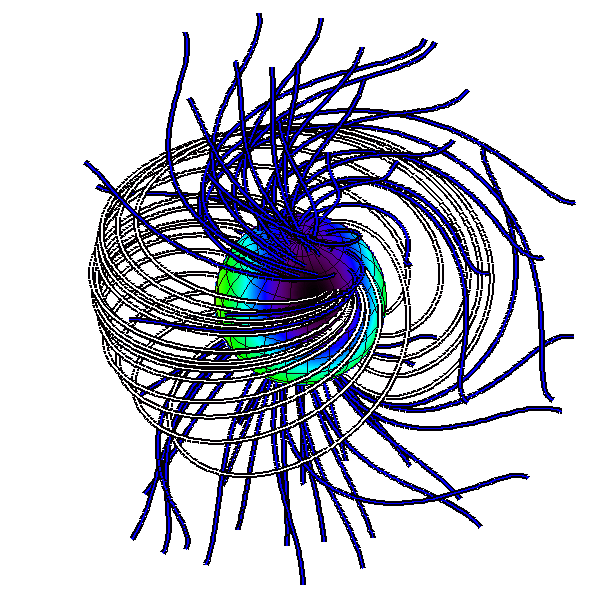}
\caption{Static field line extrapolations for CE Boo (top) and GJ 49 (bottom) for fields that are purely potential (left) and those that are the sum of potential plus non-potential (right). Closed field lines which would contain coronal gas are shown white, open field lines which would contribute to the stellar wind are shown blue.}
\label{pfss_fieldlines}
\end{figure*}


Comparing the form of our extrapolated field given in (\ref{br_pot} - \ref{bphi_pot}) with the general expressions for the observed field at the surface (\ref{br_obs} - \ref{bphi_obs}) we can see that our extrapolated field cannot match the observed surface field exactly. The reason is that the form of potential field we are using for the extrapolation (the {\it Potential Field Source Surface} method) is only one type of potential field. The assumption of a source surface forces a relationship between the field components that means they are no longer independent. While $\alpha_{lm}$  can be simply related to $B_{lm}$, we cannot match the values of $\beta_{lm}$ that are derived from the observations. Therefore this method, which selects only one type of potential field, will not be guaranteed to reproduce the potential field contribution to $B_\theta$ and $B_\phi$ that is fitted to the data. 

With this caveat in mind, we use the observed $B_r$ at the stellar surface to determine $B_{lm}$ and hence to obtain  the potential contribution to the azimuthal and meridional fields $B_\phi^{pot}$ and $B_\theta^{pot}$. We show these in the bottom rows of Fig. \ref{CEBoo_static} and Fig. \ref{GJ49_static}. It is clear by comparison with the observed surface maps shown in Figs. \ref{observations_CEBoo} and \ref{observations_GJ49}, that this potential field does not reproduce all the observed field components. In particular, the uni-directional band of azimuthal field is absent from these potential field maps. In order to extrapolate the non-potential part of the field, however, we need to make an assumption about the nature of the coronal currents. We base our extrapolation on the method developed by \citet{hussain_current_02}. This is not a force-free solution, but it allows us to incorporate fully the non-potential contribution of the surface field and to extrapolate it into the corona.

\subsection{Non-potential field extrapolation}
\label{non_potential_section}
In general, the magnetic field will be a sum of potential and non-potential terms such that $\underline{B}=\underline{B}^{pot} + \underline{B}^{np}$. We assume that the non-potential magnetic field is perpendicular to the radial direction (i.e. it lies on spherical shells and so $B_r^{np} =  0$).  Furthermore, the electric currents are assumed to be derived from a potential $Q$:
\begin{equation}
\nabla \times {\bf B}^{\rm np} = - \nabla Q . 
\end{equation}
It follows that $\nabla^2 Q = 0$, so $Q({\bf r})$ has a solution in terms of spherical harmonics. As shown in the Appendix, we find solutions for this non-potential magnetic field that vanish at the source surface and have the form 
\begin{equation}
B_r^{np} =  0 
\end{equation}
\begin{equation}
B_\theta^{np}   =   -   \sum^N_{l=1}\sum^l_{m=-l} 
           l(l+1) C_{lm} \frac{P_{lm}(\theta)}{\sin\theta}imh_l(r,R_s)r^{-(l+1)}    e^{im\phi}
\label{btheta_np}
\end{equation}
\begin{equation}
B_\phi^{np}   =   \sum^N_{l=1}\sum^l_{m=-l} 
                l(l+1) C_{lm}\frac{dP_{lm}(\theta)}{d\theta} h_l(r,R_s)r^{-(l+1)}  e^{im\phi}   
\label{bphi_np}       
\end{equation}
where
\begin{equation}
 h_l(r,R_s) = \left[ \frac{1-(r/R_s)^{2l+1}}{l+(l+1)(1/R_s)^{2l+1}} \right]
\end{equation}
and as $R_s\rightarrow \infty$ we recover $h_l(1)\rightarrow 1/l$.

While this is not the most general form of non-potential field, it has the useful property that the equations for $B^{pot}$ and $B^{np}$ are now structurally very similar to the forms used in (\ref{br_obs} - \ref{bphi_obs}) to describe the surface field. The coefficients $\gamma_{lm}$ and $C_{lm}$ that govern the non-potential field components can be simply related. This allows us to match the observed non-potential component of the field exactly to our model and to extrapolate it into the corona.   Thus while the potential part of our extrapolated field will not reproduce an exact match to the potential part of the observed surface field, the non-potential part matches exactly.

We therefore show the non-potential (top row) and potential (bottom row) parts of the field separately in Figs. \ref{CEBoo_static} and \ref{GJ49_static}. The total field is the sum of both of these. Fig. \ref{pfss_fieldlines} shows the extrapolation of this total field with a source surface chosen to be at $4R_\star$. The largest closed field lines have been selected in order to highlight the structure of the large-scale field. The tilt of the dipole axis can be clearly seen in both cases, although it should be noted that the rotation axes of both stars have the same inclination to the observer's line of sight. While the extrapolation of the potential contribution to the total field is fairly similar in both stars, the inclusion of the non-potential contribution highlights the differences between the magnetic field structures of the two stars. The non-potential component introduces an azimuthal shear into the field that is most apparent in GJ 49 (for which 52$\%$ of the total magnetic energy in the surface field is non-potential).



\section{The Stellar Wind Model}\label{sec.wind}

To perform the stellar wind simulations, we use the three-dimensional MHD numerical code BATS-R-US developed at University of Michigan \citep{1999JCoPh.154..284P}. 
BATS-R-US has been widely used to simulate, e.g., the Earth's magnetosphere \citep{2006AdSpR..38..263R}, the heliosphere \citep{2003ApJ...595L..57R}, the outer-heliosphere \citep{1998JGR...103.1889L, 2003ApJ...591L..61O, 2004ApJ...611..575O}, coronal mass ejections \citep{2004JGRA..10901102M,2005ApJ...627.1019L},  the magnetosphere of planets \citep{2004JGRA..10911210T,2005GeoRL..3220S06H}, and stellar winds of cool stars \citep{2009ApJ...703.1734V, 2012MNRAS.423.3285V}.
It solves the ideal MHD equations, that in the conservative form are given by
\begin{equation}
\label{eq:continuity_conserve}
\frac{\partial \rho}{\partial t} + \bf{\nabla}\cdot \left(\rho {\bf {u}}\right) = 0,
\end{equation}
\begin{equation}
\label{eq:momentum_conserve}
\frac{\partial \left(\rho {\bf {u}}\right)}{\partial t} + \bf{\nabla}\cdot\left[ \rho{\bf {u}\,\bf{u}}+ \left(p + \frac{B^2}{8\pi}\right)I - \frac{{\bf {B}\,\bf{B}}}{4\pi}\right] = \rho {\bf {g}},
\end{equation}
\begin{equation}
\label{eq:bfield_conserve}
\frac{\partial {\bf {B}}}{\partial t} + \bf{\nabla}\cdot\left({\bf {u\,B}} - {\bf {B\,u}}\right) = 0,
\end{equation}
\begin{equation}
\label{eq:energy_conserve}
\frac{\partial\varepsilon}{\partial t} +  \bf{\nabla} \cdot \left[ {\bf {u}} \left( \varepsilon + p + \frac{B^2}{8\pi} \right) - \frac{\left({\bf {u}}\cdot{\bf {B}}\right) {\bf {B}}}{4\pi}\right] = \rho {\bf {g}}\cdot {\bf {u}} ,
\end{equation}
where the eight primary variables are the mass density $\rho$, the plasma velocity ${\bf u}=\{ u_r, u_\theta, u_\varphi\}$, the magnetic field ${\bf B}=\{ B_r, B_\theta, B_\varphi\}$, and the gas pressure $p$. The gravitational acceleration due to the star with mass $M_\star$ and radius $R_\star$ is given by ${\bf g}$, and $\varepsilon$ is the total energy density given by 
\begin{equation}\label{eq:energy_density}
\varepsilon=\frac{\rho u^2}{2}+\frac{p}{\gamma-1}+\frac{B^2}{8\pi} .
\end{equation}
We consider an ideal gas, so $p=n k_B T$, where  $k_B$ is the Boltzmann constant, $T$ is the temperature, $n=\rho/(\mu m_p)$ is the particle number density of the stellar wind, $\mu m_p$ is the mean mass of the particle, and $\gamma$ is the polytropic index (such that $p \propto \rho^\gamma$).

As the initial state of the simulations, we assume that the wind is thermally driven \citep{1958ApJ...128..664P}. At the base of the corona ($r=R_\star$), we adopt a wind coronal temperature $T_0 = 2\times 10^6$~K and wind number density $n_0=10^{11}$cm$^{-3}$. The stellar rotation period $P_{\rm rot}$, $M_\star$ and $R_\star$ are given in Table \ref{parameters}. 
 With this numerical setting, the initial solution for the density, pressure (or temperature) and wind velocity profiles are fully specified. 

To complete our initial numerical set up, we assume that the magnetic field is either potential everywhere (i.e., $\bf\nabla \times {\bf B}=0$) or the sum of potential plus non-potential components, as described in sections (\ref{potential_section}) and (\ref{non_potential_section}). The initial solution for ${\bf B}$ is found once the distance to the source surface is assumed (set at $4~R_\star$ in the initial state of our runs) and the surface magnetic field is specified: either simply the radial component (in the case of a potential field) or all three components (in the case of a total potential plus non-potential field). 

Once set at the initial state of the simulation, the distribution of $B_r$ is held fixed at the surface of the star throughout the simulation run, as are the coronal base density and thermal pressure. A zero radial gradient is set to the remaining components of ${\bf B}$ and ${\bf u}=0$ in the frame corotating with the star. The outer boundaries at the edges of the grid have outflow conditions, i.e., a zero gradient is set to all the primary variables. The rotation axis of the star is aligned with the $z$-axis, and the star is assumed to rotate as a solid body.

Our grid is Cartesian and extends in $x$, $y$, and $z$ from $-20$ to $20~R_\star$, with the star placed at the origin of the grid. BATS-R-US uses block adaptive mesh refinement, which allows for variation in numerical resolution within the computational domain. The finest resolved cells are located close to the star (for $r \lesssim 2~R_\star$), where the linear size of the cubic cell is $0.02~R_\star$. The coarsest cell is about one order of magnitude larger (linear size of $0.31~R_\star$) and is located at the outer edges of the grid. The total number of cells in our simulations is about $15$ million. 

As the simulations evolve in time, both the wind and magnetic field lines are allowed to interact with each other. The resultant solution, obtained self-consistently, is found when the system reaches a steady state (in the reference frame corotating with the star).  




\begin{figure}
		\includegraphics[width=80mm]{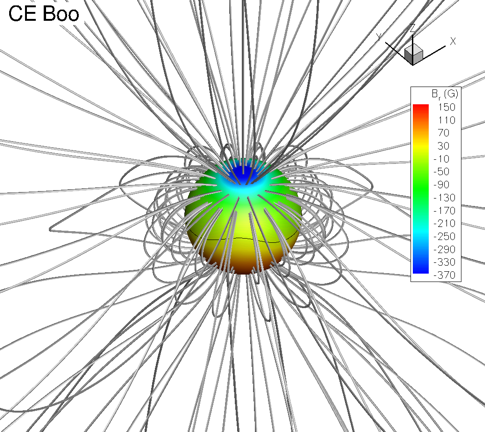}
		\includegraphics[width=80mm]{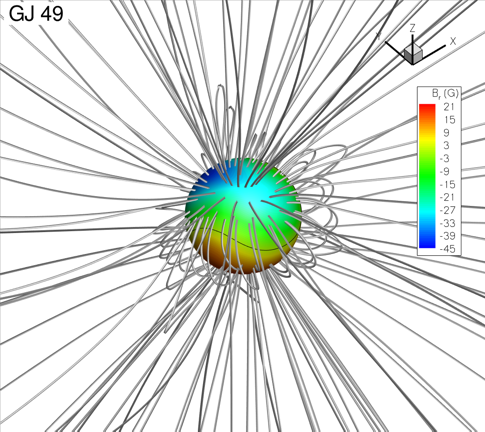}
		\includegraphics[width=80mm]{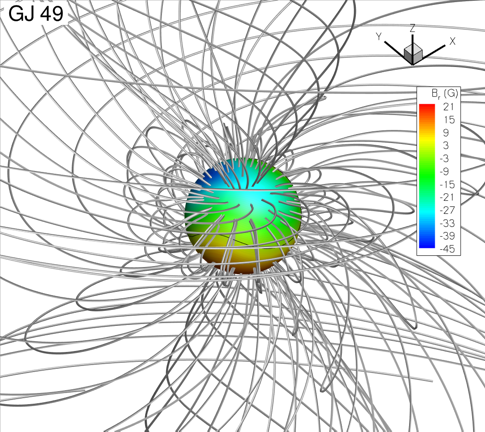}
\caption{Final magnetic field structures for CE Boo (top row) and GJ 49 (middle row). The final state is the same, regardless of whether the initial state is the total field (i.e. the potential plus non-potential field), or simply the potential field. The bottom row shows the effect on GJ 49 of artificially decreasing the rotation period from 18.6 to 0.6 days}
\label{Aline}
\end{figure}

\section{Results and Discussion}
We have separated the magnetic fields of CE Boo and GJ 49 into their lowest-energy (potential) and stressed (non-potential) components. This has allowed us to isolate both the locations where the field is stressed above its lowest energy state and also the nature of the structures that carry these stresses. We find that the departures from a lowest-energy state are apparent mainly in the azimuthal field (the meridional field contributes a negligibly-small non-potential component). This appears as a clearly-defined mid-latitude band of unidirectional azimuthal field (see Figs. \ref{CEBoo_static} and \ref{GJ49_static}). This is similar to the non-potential field of the young rapid rotator AB Dor \citep{hussain_current_02}  except that it appears at lower latitudes. 

By extrapolating these surface fields into the corona we can see that the presence of the non-potential field does not change the overall topology of the coronal field, but it provides an azimuthal shear (see Fig. \ref{pfss_fieldlines}). We have also explored the nature of the winds that might be associated with these surface fields by using these extrapolated fields as an initial state for an MHD wind model. As the solution evolves towards a steady-state, only the radial component of the surface field is kept fixed - the azimuthal and meridional field components are allowed to vary in response to the available forces.

The top and middle rows of Fig. \ref{Aline} show the final structure of the magnetic fields of CE Boo and GJ 49. This final state was the same, regardless of whether the initial state was the total field (ie the potential plus non-potential field), or simply the potential field. This happens because it is the boundary conditions that will control the final state and the initial conditions of the system will be "flushed out" by the wind. Therefore, we find that the mass loss rates, angular momentum loss rates, and the fluxes of surface magnetic field and open magnetic field are the same in both cases. Since there are no forces in the wind model capable of providing the stresses necessary to sustain the strong azimuthal field at the stellar surface, the solution relaxes back to something close to a potential field at the surface. As a result, the steady-state wind solution has meridional and azimuthal field components at the stellar surface that are close to the potential component of the field that we calculate and that do not reproduce the observed non-potential component of the surface field.

At larger heights of course, the action of the wind stresses the field and generates an azimuthal field component, but this is fairly small at the stellar surface, particularly for such slowly-rotating stars. In order to confirm the role of rotation in influencing the field structure, we also artificially increase the rotation rate of GJ 49, while keeping the initial magnetic field structure unchanged. The resulting field structure is shown in Fig. \ref{Aline}. While an azimuthal field develops with height in the corona, it is small at the surface and cannot explain the observations. This suggests that this azimuthal field is produced not by the wind, but by the sub-surface dynamo. 


These simulations therefore suggest that the ambient winds of these slowly-rotating stars are well described by the potential components of their surface fields. The strong azimuthal fields seen at the surface should not survive to the heights in the corona at which the wind is launched. They may of course be important in determining flare locations and energies. For GJ 49, for example, $52\%$ of the total magnetic energy close to the surface is contained in the non-potential part of the field and is therefore available for release. It is mainly contained in a well-defined band that is centred around latitudes 30$^\circ$ - 40$^\circ$. This is the maximum latitude at which solar active regions are seen and from which solar coronal mass ejections are launched. This might suggest that this is the region from which flares and coronal mass ejections could be expected. On the young rapid rotator  AB Dor, by comparison, (P$_{\rm rot}$ = 0.514 days) the band of non-potential field is strongest around  70$^\circ$ - 80$^\circ$ \citep{hussain_current_02} which may suggest a different pattern of coronal mass ejection. Such coronal mass ejections would temporarily increase the mass loading of the stellar wind and also its ram pressure, which is responsible for compressing the magnetospheres of any orbiting planets. Whether coronal mass ejections provide a significant contribution to either angular momentum loss or the impact of the wind on orbiting planets depends on their size and frequency. The background stellar wind that we find however  is independent of the strong non-potential component of the surface fields and is primarily governed by their radial component.

\appendix
\section{The non-potential field}
We look for solutions for the non-potential field that are of the form
\begin{equation}
B_{{\rm np},r} = 0, ~~~~ B_{{\rm np},\theta} = \frac{1}{r \sin \theta} \frac{\partial F}{\partial \phi} , ~~~~ B_{{\rm np},\phi} = - \frac{1}{r} \frac{\partial F}{\partial \theta} , \label{eq:Bnp}
\end{equation}
where $F({\bf r})$ is a scalar function. These automatically satisfy $\underline{\nabla}\cdot\underline{B}=0$. Furthermore, the electric currents are assumed to be derived from a potential $Q$:
\begin{equation}
\nabla \times {\bf B}_{\rm np} = - \nabla Q . \label{eq:curlBnp}
\end{equation}
It follows that $\nabla^2 Q = 0$, so $Q({\bf r})$ has a solution in terms of spherical harmonics.
Inserting equation (\ref{eq:Bnp}) into (\ref{eq:curlBnp}), we find  from the radial component:
\begin{equation}
- \frac{1}{r^2} \left[ \frac{1}{\sin \theta} \frac{\partial} {\partial \theta} \left( \sin \theta \frac{\partial F}{\partial \theta} \right)+ \frac{1}{\sin^2 \theta} \frac{\partial^2 F} {\partial \phi^2} \right] = - \frac{\partial Q} {\partial r} ,  \label{eq:dQdr}
\end{equation}
and from the $\theta$ and $\phi$ components:
\begin{equation}
\frac{1}{r} \frac{\partial}{\partial r} \left( \frac{\partial F}{\partial \theta} \right) = - \frac{1}{r} \frac{\partial Q}{\partial \theta}  ~~~ \hbox{and} ~~~
\frac{1}{r} \frac{\partial}{\partial r} \left( \frac{1}{\sin \theta} \frac{\partial F}{\partial \phi} \right) = - \frac{1}{r \sin \theta} \frac{\partial Q}{\partial \phi} ,
\end{equation}
which can be integrated with respect to $\theta$ and $\phi$:
\begin{equation}
\frac{\partial F} {\partial r}  = - Q .  \label{eq:Q}
\end{equation}
We now introduce a third scalar $C({\bf r})$ such that
\begin{equation}
F = r^2 \frac{\partial C} {\partial r} , \label{eq:F}
\end{equation}
then equation (\ref{eq:Q}) yields
\begin{equation}
Q = - \frac{\partial}{\partial r} \left( r^2 \frac{\partial C}{\partial r} \right) .
\end{equation}
Inserting these expressions for $F$ and $Q$ into equation (\ref{eq:dQdr}) we obtain:
\begin{equation}
\frac{\partial}{\partial r} \left( r^2 \nabla^2 C \right) = 0 .
\end{equation}
Assuming $\nabla^2 C = 0$ at the stellar surface, it follows that this condition is true at all heights, so $C({\bf r})$ is also a harmonic function. We now write $C$ as a sum over spherical harmonics:
\begin{equation}
C(r,\theta,\phi) = \sum_{lm} C_{lm} q_{l} (r) P_{lm} (\theta) e^{i m \phi} ,
\end{equation}
where $l$ is the harmonic degree ($l = 1, 2, \cdots$), $m$ is the azimuthal mode number ($-l \le m \le +l$), $P_{lm} (\theta)$ is the associate Legendre function, and $q_{l} (r)$ describes the radial dependence of the various modes. Note that the function $q_{l} (r)$ defined here is different from the function $f_{l} (r)$ for the potential field (see Hussain et al.~2002). Then the non-potential components of the magnetic field follow from equation (\ref{eq:F}) and (\ref{eq:Bnp}):
\begin{eqnarray}
B_{{\rm np},\theta} & = & + \sum_{lm} C_{lm} r \frac{d q_{l}} {dr} P_{lm} (\theta) \frac{im} {\sin \theta} e^{i m \phi} , \\
B_{{\rm np},\phi} & = & - \sum_{lm} C_{lm} r \frac{d q_{l}} {dr} \frac{d P_{lm}} {d \theta} e^{i m \phi} .
\end{eqnarray}
The function $q_{l} (r)$ must satisfy the following constraints. First, we assume that $q_{l} (1) = 1$, so that $C_{lm}$ are the mode amplitudes at the stellar surface ($r = 1$). Second, since $C ({\bf r})$ is a harmonic function, $q_{l} (r)$ must be a sum of a radially decreasing term $\propto r^{-l-1}$ and an increasing term $\propto r^{l}$. Third, we require that the horizontal components of the non-potential field vanish at the source surface; this implies $dq_{l} /dr = 0$ at $r = R_{\rm s}$. From these conditions it follows that:
\begin{equation}
r \frac{d q_{l}} {dr}  =  -l(l+1)  r^{-(l+1)} h_l(r,R_s)
\end{equation}
and hence we recover expressions (\ref{btheta_np}) and (\ref{bphi_np}) for $B_\theta^{np}$ and  $B_\phi^{np}$.

\section*{Acknowledgements}
The authors acknowledge support from STFC. AAV acknowledges support from the Royal Astronomical Society. JM acknowledges 
support from a fellowship of the Alexander von Humboldt foundation.


\begin{thebibliography}{43}
\expandafter\ifx\csname natexlab\endcsname\relax\def\natexlab#1{#1}\fi

\bibitem[{{Aarnio} {et~al.}(2011{\natexlab{a}}){Aarnio}, {Stassun}, {Hughes},
  \& {McGregor}}]{aarnio_solarCMEs_11}
{Aarnio} A.~N., {Stassun} K.~G., {Hughes} W.~J., {McGregor} S.~L.,
  2011{\natexlab{a}}, \solphys, 268, 195

\bibitem[{{Aarnio} {et~al.}(2011{\natexlab{b}}){Aarnio}, {Stassun}, {Matt},
  {Hughes}, \& {McGregor}}]{aarnio_cs16_11}
{Aarnio} A.~N., {Stassun} K.~G., {Matt} S.~P., {Hughes} W.~J., {McGregor}
  S.~L., 2011{\natexlab{b}}, in Astronomical Society of the Pacific Conference
  Series, Vol. 448, 16th Cambridge Workshop on Cool Stars, Stellar Systems, and
  the Sun, {Johns-Krull} C., {Browning} M.~K., {West} A.~A., eds., p.~43

\bibitem[{{Altschuler} \& {Newkirk, Jr.}(1969)}]{altschuler69}
{Altschuler} M.~D., {Newkirk, Jr.} G., 1969, \sp, 9, 131

\bibitem[{Barnes {et~al.}(2000)Barnes, Collier~Cameron, James, \&
  Donati}]{barnes20PZTel}
Barnes J., Collier~Cameron A., James D.~J., Donati J.-F., 2000, \mn, 314, 162

\bibitem[{{Barnes} {et~al.}(2005){Barnes}, {Collier Cameron}, {Donati},
  {James}, {Marsden}, \& {Petit}}]{barnes_DR_05}
{Barnes} J.~R., {Collier Cameron} A., {Donati} J.-F., {James} D.~J., {Marsden}
  S.~C., {Petit} P., 2005, \mn, 357, L1

\bibitem[{{Brickhouse} \& {Dupree}(1998)}]{brickhouse98}
{Brickhouse} N., {Dupree} A., 1998, \apj, 502, 918

\bibitem[{{Byrne} {et~al.}(1996){Byrne}, {Eibe}, \& {Rolleston}}]{byrne96hkaqr}
{Byrne} P., {Eibe} M., {Rolleston} W., 1996, \aanda, 311, 651

\bibitem[{Collier~Cameron \& Robinson(1989{\natexlab{a}})}]{cameron89eject}
Collier~Cameron A., Robinson R.~D., 1989{\natexlab{a}}, \mn, 238, 657

\bibitem[{Collier~Cameron \& Robinson(1989{\natexlab{b}})}]{cameron89cloud}
---, 1989{\natexlab{b}}, \mn, 236, 57

\bibitem[{Collier~Cameron \& Woods(1992)}]{cameron92alpper}
Collier~Cameron A., Woods J.~A., 1992, \mn, 258, 360

\bibitem[{{Donati} {et~al.}(2008){Donati}, {Morin}, {Petit}, {Delfosse},
  {Forveille}, {Auri{\`e}re}, {Cabanac}, {Dintrans}, {Fares}, {Gastine},
  {Jardine}, {Ligni{\`e}res}, {Paletou}, {Velez}, \&
  {Th{\'e}ado}}]{morin_earlyM_08}
{Donati} J., {Morin} J., {Petit} P., {Delfosse} X., {Forveille} T.,
  {Auri{\`e}re} M., {Cabanac} R., {Dintrans} B., {Fares} R., {Gastine} T.,
  {Jardine} M.~M., {Ligni{\`e}res} F., {Paletou} F., {Velez} J.~C.~R.,
  {Th{\'e}ado} S., 2008, \mnras, 390, 545

\bibitem[{Donati \& Brown(1997)}]{donati97}
Donati J.-F., Brown S., 1997, \aanda, 326, 1135

\bibitem[{Donati \& Collier~Cameron(1997)}]{donati97abdor95}
Donati J.-F., Collier~Cameron A., 1997, \mnras, 291, 1

\bibitem[{{Donati} {et~al.}(2006){Donati}, {Howarth}, {Jardine}, {Petit},
  {Catala}, {Landstreet}, {Bouret}, {Alecian}, {Barnes}, {Forveille},
  {Paletou}, \& {Manset}}]{donati06tausco}
{Donati} J.-F., {Howarth} I.~D., {Jardine} M.~M., {Petit} P., {Catala} C.,
  {Landstreet} J.~D., {Bouret} J.-C., {Alecian} E., {Barnes} J.~R., {Forveille}
  T., {Paletou} F., {Manset} N., 2006, \mnras, 370, 629

\bibitem[{Donati {et~al.}(2000)Donati, Mengel, Carter, Cameron, \&
  Wichmann}]{donati20RXJ}
Donati J.-F., Mengel M., Carter B., Cameron A., Wichmann R., 2000, \mn, 316,
  699

\bibitem[{{Dupree} {et~al.}(1993){Dupree}, {Brickhouse}, {Doschek}, {Green}, \&
  {Raymond}}]{dupree93}
{Dupree} A., {Brickhouse} N., {Doschek} G., {Green} J., {Raymond} J., 1993,
  \apj, 418, L41

\bibitem[{{Eibe}(1998)}]{eibe98re1816}
{Eibe} M.~T., 1998, \aanda, 337, 757

\bibitem[{{G{\"u}del}(2004)}]{guedel_review_04}
{G{\"u}del} M., 2004, \aandar, 12, 71

\bibitem[{{G\"udel} {et~al.}(2001){G\"udel}, {Audard}, {den Boggende},
  {Brinkman}, {den Herder}, J.S., {Mewe}, {Raassen}, {de Vries}, {Behar},
  {Cottam}, {Kahn}, {Paerels}, {Peterson}, {Rasmussen}, {Sako},
  {Branduardi-Raymont}, {Sakelliou}, \& {Erd}}]{gudel01}
{G\"udel} M., {Audard} M., {den Boggende} A., {Brinkman} A., {den Herder} J.,
  J.S. K., {Mewe} R., {Raassen} A., {de Vries} C., {Behar} E., {Cottam} J.,
  {Kahn} S., {Paerels} F., {Peterson} J., {Rasmussen} A., {Sako} M.,
  {Branduardi-Raymont} G., {Sakelliou} I., {Erd} C., 2001, in Proceedings of
  ``X-ray astronomy 2000'', R.~Giaconni L.~Stella S.~S., ed., ASP conference
  series
  
  \bibitem[{{Hansen} {et~al.}(2005){Hansen}, {Ridley}, {Hospodarsky},
  {Achilleos}, {Dougherty}, {Gombosi}, \& {T{\'o}th}}]{2005GeoRL..3220S06H}
{Hansen} K.~C., {Ridley} A.~J., {Hospodarsky} G.~B., {Achilleos} N.,
  {Dougherty} M.~K., {Gombosi} T.~I., {T{\'o}th} G., 2005, \grl, 32, 20


\bibitem[{{Hussain} {et~al.}(2002){Hussain}, {van Ballegooijen}, {Jardine}, \&
  {Collier Cameron}}]{hussain_current_02}
{Hussain} G.~A.~J., {van Ballegooijen} A.~A., {Jardine} M., {Collier Cameron}
  A., 2002, \apj, 575, 1078

\bibitem[{{Jardine} {et~al.}(2002){Jardine}, {Collier Cameron}, \&
  {Donati}}]{jardine02structure}
{Jardine} M., {Collier Cameron} A., {Donati} J.-F., 2002, \mn, 333, 339

\bibitem[{{Jardine} \& {van Ballegooijen}(2005)}]{jardine05proms}
{Jardine} M., {van Ballegooijen} A.~A., 2005, \mnras, 361, 1173

\bibitem[{{Jeffers} {et~al.}(2011){Jeffers}, {Donati}, {Alecian}, \&
  {Marsden}}]{jeffers_HD171488_11}
{Jeffers} S.~V., {Donati} J.-F., {Alecian} E., {Marsden} S.~C., 2011, \mn, 411,
  1301

\bibitem[{{Jeffries}(1993)}]{jeffries93}
{Jeffries} R., 1993, \mn, 262, 369

\bibitem[{{Khodachenko} {et~al.}(2007){Khodachenko}, {Ribas}, {Lammer},
  {Grie{\ss}meier}, {Leitner}, {Selsis}, {Eiroa}, {Hanslmeier}, {Biernat},
  {Farrugia}, \& {Rucker}}]{khodachenko_CME_07}
{Khodachenko} M.~L., {Ribas} I., {Lammer} H., {Grie{\ss}meier} J.-M., {Leitner}
  M., {Selsis} F., {Eiroa} C., {Hanslmeier} A., {Biernat} H.~K., {Farrugia}
  C.~J., {Rucker} H.~O., 2007, Astrobiology, 7, 167

\bibitem[{{Linde} {et~al.}(1998){Linde}, {Gombosi}, {Roe}, {Powell}, \&
  {Dezeeuw}}]{1998JGR...103.1889L}
{Linde} T.~J., {Gombosi} T.~I., {Roe} P.~L., {Powell} K.~G., {Dezeeuw} D.~L.,
  1998, \jgr, 103, 1889

\bibitem[{{Lugaz} {et~al.}(2005){Lugaz}, {Manchester}, \&
  {Gombosi}}]{2005ApJ...627.1019L}
{Lugaz} N., {Manchester} IV W.~B., {Gombosi} T.~I., 2005, \apj, 627, 1019

\bibitem[{{Maggio} {et~al.}(2000){Maggio}, {Pallavicini}, {Reale}, \&
  {Tagliaferri}}]{maggio2000}
{Maggio} A., {Pallavicini} R., {Reale} F., {Tagliaferri} G., 2000, \aanda, 356,
  627

\bibitem[{{Manchester} {et~al.}(2004){Manchester}, {Gombosi}, {Roussev}, {De
  Zeeuw}, {Sokolov}, {Powell}, {T{\'o}th}, \& {Opher}}]{2004JGRA..10901102M}
{Manchester} W.~B., {Gombosi} T.~I., {Roussev} I., {De Zeeuw} D.~L., {Sokolov}
  I.~V., {Powell} K.~G., {T{\'o}th} G., {Opher} M., 2004, Journal of
  Geophysical Research (Space Physics), 109, 1102

\bibitem[{{Marsden} {et~al.}(2006){Marsden}, {Donati}, {Semel}, {Petit}, \&
  {Carter}}]{marsden_DR_06}
{Marsden} S.~C., {Donati} J.-F., {Semel} M., {Petit} P., {Carter} B.~D., 2006,
  \mn, 370, 468

\bibitem[{{Marsden} {et~al.}(2005){Marsden}, {Waite}, {Carter}, \&
  {Donati}}]{marsden_DR_05}
{Marsden} S.~C., {Waite} I.~A., {Carter} B.~D., {Donati} J.-F., 2005, \mn, 359,
  711

\bibitem[{{Mestel}(1999)}]{mestel_book_99}
{Mestel} L., 1999, {Stellar magnetism}

\bibitem[{{Morin}(2012)}]{morin12}
{Morin} J., 2012, in EAS Publications Series, Vol.~57, EAS Publications Series,
  {Reyl{\'e}} C., {Charbonnel} C., {Schultheis} M., eds., pp. 165--191

\bibitem[{{Morin} {et~al.}(2008){Morin}, {Donati}, {Petit}, {Delfosse},
  {Forveille}, {Albert}, {Auri{\`e}re}, {Cabanac}, {Dintrans}, {Fares},
  {Gastine}, {Jardine}, {Ligni{\`e}res}, {Paletou}, {Ramirez Velez}, \&
  {Th{\'e}ado}}]{morin_midM_08}
{Morin} J., {Donati} J., {Petit} P., {Delfosse} X., {Forveille} T., {Albert}
  L., {Auri{\`e}re} M., {Cabanac} R., {Dintrans} B., {Fares} R., {Gastine} T.,
  {Jardine} M.~M., {Ligni{\`e}res} F., {Paletou} F., {Ramirez Velez} J.~C.,
  {Th{\'e}ado} S., 2008, \mnras, 390, 567

\bibitem[{{Morin} {et~al.}(2010){Morin}, {Donati}, {Petit}, {Delfosse},
  {Forveille}, \& {Jardine}}]{morin_lateM_10}
{Morin} J., {Donati} J., {Petit} P., {Delfosse} X., {Forveille} T., {Jardine}
  M.~M., 2010, \mnras, 1077

\bibitem[{{Opher} {et~al.}(2003){Opher}, {Liewer}, {Gombosi}, {Manchester},
  {DeZeeuw}, {Sokolov}, \& {Toth}}]{2003ApJ...591L..61O}
{Opher} M., {Liewer} P.~C., {Gombosi} T.~I., {Manchester} W., {DeZeeuw} D.~L.,
  {Sokolov} I., {Toth} G., 2003, \apjl, 591, L61

\bibitem[{{Opher} {et~al.}(2004){Opher}, {Liewer}, {Velli}, {Bettarini},
  {Gombosi}, {Manchester}, {DeZeeuw}, {Toth}, \&
  {Sokolov}}]{2004ApJ...611..575O}
{Opher} M., {Liewer} P.~C., {Velli} M., {Bettarini} L., {Gombosi} T.~I.,
  {Manchester} W., {DeZeeuw} D.~L., {Toth} G., {Sokolov} I., 2004, \apj, 611,
  575

\bibitem[{{Parker}(1958)}]{1958ApJ...128..664P}
{Parker} E.~N., 1958, \apj, 128, 664

\bibitem[{{Petit} {et~al.}(2008){Petit}, {Dintrans}, {Solanki}, {Donati},
  {Auri{\`e}re}, {Ligni{\`e}res}, {Morin}, {Paletou}, {Ramirez Velez},
  {Catala}, \& {Fares}}]{petit_toroidal_08}
{Petit} P., {Dintrans} B., {Solanki} S.~K., {Donati} J.-F., {Auri{\`e}re} M.,
  {Ligni{\`e}res} F., {Morin} J., {Paletou} F., {Ramirez Velez} J., {Catala}
  C., {Fares} R., 2008, \mnras, 388, 80

\bibitem[{{Pinto} {et~al.}(2011){Pinto}, {Brun}, {Jouve}, \&
  {Grappin}}]{pinto_solarcycle_11}
{Pinto} R.~F., {Brun} A.~S., {Jouve} L., {Grappin} R., 2011, apj, 737, 72

\bibitem[{{Pointer} {et~al.}(2002){Pointer}, {Jardine}, {Collier Cameron}, \&
  {Donati}}]{pointer01evol}
{Pointer} G.~R., {Jardine} M., {Collier Cameron} A., {Donati} J.-F., 2002, \mn,
  330, 160+

\bibitem[{{Powell} {et~al.}(1999){Powell}, {Roe}, {Linde}, {Gombosi}, \& {de
  Zeeuw}}]{1999JCoPh.154..284P}
{Powell} K.~G., {Roe} P.~L., {Linde} T.~J., {Gombosi} T.~I., {de Zeeuw} D.~L.,
  1999, Journal of Computational Physics, 154, 284

\bibitem[{{Ridley} {et~al.}(2006){Ridley}, {de Zeeuw}, {Manchester}, \&
  {Hansen}}]{2006AdSpR..38..263R}
{Ridley} A.~J., {de Zeeuw} D.~L., {Manchester} W.~B., {Hansen} K.~C., 2006,
  Advances in Space Research, 38, 263

\bibitem[{{Riley} {et~al.}(2006){Riley}, {Linker}, {Miki{\'c}}, {Lionello},
  {Ledvina}, \& {Luhmann}}]{riley_PFSS_06}
{Riley} P., {Linker} J.~A., {Miki{\'c}} Z., {Lionello} R., {Ledvina} S.~A.,
  {Luhmann} J.~G., 2006, \apj, 653, 1510

\bibitem[{{Roussev} {et~al.}(2003){Roussev}, {Gombosi}, {Sokolov}, {Velli},
  {Manchester}, {DeZeeuw}, {Liewer}, {T{\'o}th}, \&
  {Luhmann}}]{2003ApJ...595L..57R}
{Roussev} I.~I., {Gombosi} T.~I., {Sokolov} I.~V., {Velli} M., {Manchester} IV
  W., {DeZeeuw} D.~L., {Liewer} P., {T{\'o}th} G., {Luhmann} J., 2003, \apjl,
  595, L57

\bibitem[{{Sanz-Forcada} {et~al.}(2003){Sanz-Forcada}, {Maggio}, \&
  {Micela}}]{sanz_forcada_abdor_03}
{Sanz-Forcada} J., {Maggio} A., {Micela} G., 2003, \aanda, 408, 1087

\bibitem[{Schatten {et~al.}(1969)Schatten, Wilcox, \& Ness}]{schatten69}
Schatten K., Wilcox J., Ness N., 1969, \sp, 6, 442

\bibitem[{{Schrijver} {et~al.}(1995){Schrijver}, {Mewe}, {van den Oord}, \&
  {Kaastra}}]{schrijver95}
{Schrijver} C., {Mewe} R., {van den Oord} G., {Kaastra} J., 1995, \aanda, 302,
  438

\bibitem[{{Strassmeier}(2009)}]{strassmeier_review_09}
{Strassmeier} K.~G., 2009, \aapr, 17, 251

\bibitem[{{T{\'o}th} {et~al.}(2004){T{\'o}th}, {Kov{\'a}cs}, {Hansen}, \&
  {Gombosi}}]{2004JGRA..10911210T}
{T{\'o}th} G., {Kov{\'a}cs} D., {Hansen} K.~C., {Gombosi} T.~I., 2004, Journal
  of Geophysical Research (Space Physics), 109, 11210

\bibitem[{van Ballegooijen {et~al.}(1998)van Ballegooijen, Cartledge, \&
  Priest}]{vanballegooijen98}
van Ballegooijen A., Cartledge N., Priest E., 1998, \apj, 501, 866

\bibitem[{{Vidotto} {et~al.}(2012){Vidotto}, {Fares}, {Jardine}, {Donati},
  {Opher}, {Moutou}, {Catala}, \& {Gombosi}}]{2012MNRAS.423.3285V}
{Vidotto} A.~A., {Fares} R., {Jardine} M., {Donati} J.-F., {Opher} M., {Moutou}
  C., {Catala} C., {Gombosi} T.~I., 2012, \mnras, 423, 3285

\bibitem[{{Vidotto} {et~al.}(2009){Vidotto}, {Opher}, {Jatenco-Pereira}, \&
  {Gombosi}}]{2009ApJ...703.1734V}
{Vidotto} A.~A., {Opher} M., {Jatenco-Pereira} V., {Gombosi} T.~I., 2009, \apj,
  703, 1734

\bibitem[{{Wang} {et~al.}(2006){Wang}, {Sheeley}, \&
  {Rouillard}}]{wang_nonaxisym_11}
{Wang} Y.-M., {Sheeley} Jr. N.~R., {Rouillard} A.~P., 2006, \apj, 644, 638

\bibitem[{{Weber} \& {Davies}(1967)}]{weber67}
{Weber} E., {Davies} L., 1967, \apj, 148, 217

\end{thebibliography}

\end{document}